\documentclass[lettersize,journal]{IEEEtran}
\usepackage{amsmath,amsfonts}
\usepackage{algorithmic}
\usepackage{algorithm}
\usepackage{array}
\usepackage[caption=false,font=normalsize,labelfont=sf,textfont=sf]{subfig}
\usepackage{textcomp}
\usepackage{stfloats}
\usepackage{url}
\usepackage{verbatim}
\usepackage{graphicx}
\usepackage{cleveref}
\usepackage{cite}

\usepackage{tabu}                      % only used for the table example
\usepackage{booktabs}                  % only used for the table example
\usepackage{lipsum}                    % used to generate placeholder text
\usepackage{mwe}                       % used to generate placeholder figures

\usepackage{mathptmx}                  % use matching math font
\usepackage{amsmath}
\usepackage{amsfonts}
\usepackage{tabularx}
\usepackage{multirow} 

\usepackage{xcolor}
\usepackage{colortbl}
\usepackage{afterpage}                % for precise table placement
\usepackage{wrapfig}                  % for text wrapping around figures
\usepackage{caption}
\usepackage{svg}

\begin{document}

\title{SEGA: Drivable 3D Gaussian Head Avatar from a Single Image}

\author{Chen Guo$^{\ast}$, Zhuo Su$^{\ast}$, Liao Wang$^{\ast}$, Jian Wang, Shuang Li, Xu Chang, Zhaohu Li, Yang Zhao\\
Guidong Wang, Yebin Liu,~\IEEEmembership{Member,~IEEE}, Ruqi Huang,~\IEEEmembership{Member,~IEEE}
% ~\IEEEmembership{Staff,~IEEE,}
        % <-this % stops a space

% \thanks{}
% <-this % stops a space
\thanks{$^{\ast}$Chen Guo, Zhuo Su, and Liao Wang contributed equally to this work.}
\thanks{Chen Guo and Ruqi Huang are with Tsinghua Shenzhen International Graduate School, China.}
\thanks{Zhuo Su, Liao Wang, Jian Wang, Shuang Li, Xu Chang, Zhaohu Li, Yang Zhao and Guidong Wang are with ByteDance, China.}
\thanks{Chen Guo is also with ByteDance and Pengcheng Lab, China}
\thanks{Yebin Liu is with Department of Automation, Tsinghua University, China}
\thanks{Corresponding author: Zhuo Su (email: suzhuo13@gmail.com), Ruqi Huang (email: ruqihuang@sz.tsinghua.edu.cn).}
}

\markboth{Journal of \LaTeX\ Class Files,~Vol.~14, No.~8, August~2021}%
{Shell \MakeLowercase{\textit{et al.}}: A Sample Article Using IEEEtran.cls for IEEE Journals}


\let\oldtwocolumn\twocolumn
\renewcommand\twocolumn[1][]{%
        \oldtwocolumn[{#1}{
                \begin{center}
                \vspace{-8mm}
                        \includegraphics[width=\linewidth]{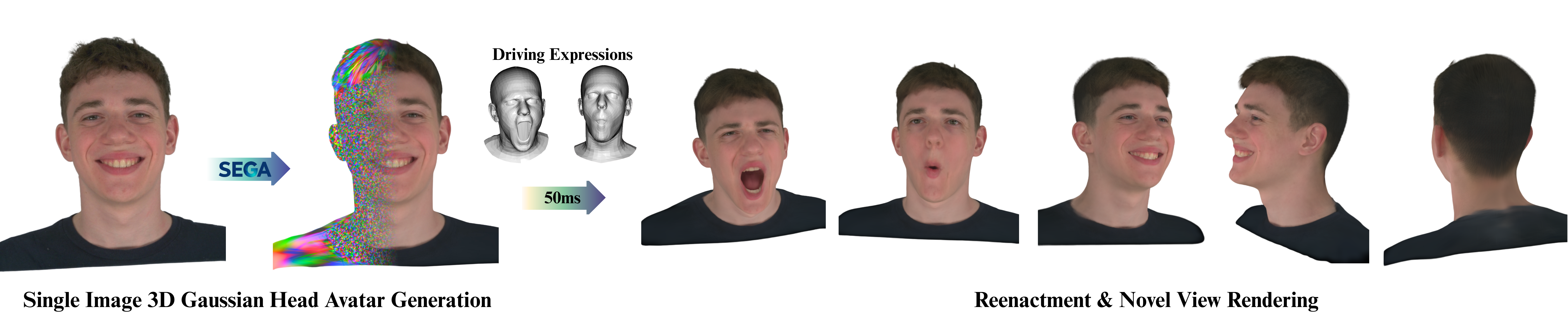}
                        \captionof{figure}{We introduce \textbf{SEGA}, a novel approach for reconstructing photorealistic 3D Gaussian splats of a human head from a single image. Once the avatar is generated, SEGA enables 360-degree free-viewpoint rendering, as well as self-identity and cross-identity reenactment in real-time.} 
                        \label{fig:teaser}
                \end{center}
        }]
}
% === Revision color toggle ===
% \revisiontrue  → blue highlights (revision mode)
% \revisionfalse → pure black (final version)
\newif\ifrevision
% \revisiontrue  % ← Change to \revisionfalse for final black version
\revisionfalse
\ifrevision
  \colorlet{revcolor}{blue!70!black}
\else
  \colorlet{revcolor}{black}
\fi

\definecolor{revised}{rgb}{0.1,0.1,0.74}
\newcommand{\revised}[1]{\color{revised}}

\maketitle

\begin{abstract}
Creating photorealistic 3D head avatars from limited input has become increasingly important for applications in virtual reality, telepresence, and digital entertainment. While recent advances like neural rendering and 3D Gaussian Splatting have enabled high-quality digital human avatar creation and animation, 
% most methods require multiple images or multi-view inputs to achieve high-fidelity results, limiting their practicality in real-world scenarios.
most methods rely on multiple images or multi-view inputs, limiting their practicality for real-world use.
In this paper, we propose \textbf{SEGA}, a novel approach for \textbf{S}ingle-imag\textbf{E}-based 3D drivable \textbf{G}aussian head \textbf{A}vatar creation which enables full 360-degree head rendering.  \textbf{SEGA}
builds on two key insights: (1) a hierarchical static--dynamic decomposition, and (2) the integration of 2D vision priors with 3D data. Our Static Branch leverages a large reconstruction model with FLAME priors to capture rigid, expression-invariant regions (e.g., forehead, scalp), ensuring strong identity preservation and viewpoint generalization. 
The Dynamic Branch adopts a lightweight VQ-VAE to model deformable regions (e.g., mouth, eyes, cheeks), enabling real-time synthesis of fine-grained expression details. 
SEGA further fuses 2D vision priors (DINOv2 and a pretrained CodeFormer encoder) with limited multi-view, multi-expression, and multi-identity 3D data. 
This design ensures robust generalization to unseen identities while preserving 3D consistency across novel viewpoints and expressions, thereby maximizing the utility of scarce 3D supervision. Furthermore, SEGA performs person-specific fine-tuning to further enhance fidelity and realism of the generated avatars.
Experiments show our method outperforms state-of-the-art approaches in generalization ability, identity preservation, and expression realism, advancing one-shot avatar creation for practical applications.
\end{abstract}


\section{Introduction}\label{sec:intro}

%% 1. Top-view
% Application
\IEEEPARstart{T}{he} creation of photorealistic 3D face avatars holds immense value for applications like virtual reality, telepresence, and digital entertainment~\cite{ma2021pixel,lombardi2021mixture,cao2022authentic, zheng2024headgap, saito2024rgca, martinez2024codec}.
% \jw{cite more papers here.}
% 3D gaussian splatting for human avatar creation.
In particular, due to the high efficiency and high rendering quality, 3D Gaussian Splatting~\cite{kerbl20233d} has been widely adopted for the creation of photo-realistic 3D avatars~\cite{saito2024rgca, qian2024gaussianavatars, xiang2024flashavatar}.
Unfortunately, these methods usually require video sequences or even calibrated multi-view images as input, which is tedious or impossible to capture and process for the general user.
Among the various input options, a single image is the most accessible and user-friendly, making it the ideal choice for widespread adoption. Nevertheless, generating high-fidelity 3D avatars from a single image remains a challenging task due to the inherently ill-posed nature of the problem. It requires inferring complex 3D geometry and texture information from limited 2D observations, which often leads to ambiguities in the depth, occlusions, and fine details.

% Among the various input options, a single image is the most accessible and user-friendly, making it the ideal choice for widespread adoption. However, generating high-fidelity 3D avatars from a single image remains a challenging task due to the inherently ill-posed nature of the problem. It requires inferring complex 3D geometry and texture information from limited 2D observations, which often leads to ambiguities in the depth, occlusions, and fine details that must be plausibly reconstructed.

Recently, several methods have been proposed to generate 3D avatars from a single image or sparse-view images, which can be broadly categorized into three approaches. First, 2D-driven methods such as GPAvatar~\cite{chu2024gpavatar}, GAGAvatar~\cite{chu2024generalizable}, Portrait4D~\cite{deng2024portrait4d}, and Portrait4Dv2~\cite{deng2024portrait4dv2} leverage large-scale 2D datasets to enhance visual fidelity and improve generalization across diverse identities. These methods excel in identity diversity but often suffer from 3D consistency issues when rendered from novel viewpoints. Second, 3D-prior-based approaches like HeadGAP~\cite{zheng2024headgap} and One2Avatar~\cite{yu2024one2avatar} incorporate generalizable 3D priors to achieve high-quality results with better geometric consistency, but their generalization is limited by the identity diversity in 3D training datasets. Third, recent foundation model-based methods such as Avat3r~\cite{kirschstein2025avat3rlargeanimatablegaussian} combine Vision Transformer backbones with multi-view generation techniques, and FaceLift~\cite{lyu2025faceliftlearninggeneralizablesingle} achieves 360-degree reconstruction through GS-LRM. \textcolor{revcolor}{Meanwhile, 4D portrait avatar methods such as CAP4D~\cite{taubner2025cap4d} and FaceCraft4D~\cite{yin2025facecraft4d} extend avatar generation into the temporal domain. UniGAHA~\cite{teotia2025unigaha} builds a universal head avatar prior that disentangles identity and expression latent spaces for audio-driven animation.} Yet these methods either require multiple temporal frames or involve complex multi-view generation processes. 
% TalkingGaussian~\cite{li2024talkinggaussianstructurepersistent3dtalking} addresses face-mouth motion inconsistency by decomposing the model into facial and intra-mouth regions to improve mouth motion and structure reconstruction. However, this binary decomposition approach is relatively coarse compared to our fine-grained static-dynamic decomposition that captures the intrinsic nature of expression-dependent variations. 
TalkingGaussian~\cite{li2024talkinggaussianstructurepersistent3dtalking} employs a deformation-based framework with Face-Mouth Decomposition to address motion inconsistencies. However, it treats all facial regions uniformly without distinguishing between expression-invariant and expression-dependent areas—a limitation our hierarchical static-dynamic decomposition addresses.
%最近，LAMintroduces a large-scale avatar model，同时使用了2D数据和3D数据进行训练。然而他总体来说还是一个静态模型，驱动的时候未能考虑表情相关的区域变化。
LAM~\cite{he2025LAM} introduces a large-scale avatar model trained with both 2D and 3D data. However, it essentially remains a static model and fails to account for expression-dependent variations during animation.

In a nutshell, the core challenge of single-image avatar generation lies in effectively bridging the gap between  2D identity diversity and 3D geometric consistency within a unified framework. Approaches relying primarily on 2D datasets often fail to preserve 3D consistency under novel viewpoints and facial expressions, while methods based on 3D datasets---though geometrically accurate---suffer from limited identity variation and thus generalize poorly to unseen subjects. As a result, current techniques still fall short of simultaneously fulfilling three critical requirements: generalization to novel views, robust expression animation, and high identity diversity.

To address these challenges, we propose SEGA, designed to meet all three requirements through two key insights: (1) a hierarchical static--dynamic decomposition, and (2) the integration of 2D vision priors with 3D data. 

% \noindent\textbf{Hierarchical Static--Dynamic Decomposition.}  
In our disentanglement strategy, the Static Branch provides strong generalization to novel viewpoints and faithful identity preservation, while the Dynamic Branch enables high-fidelity expression-driven animation while maintaining real-time performance. 
Concretely, the Static Branch employs a large reconstruction model to predict position offsets relative to the standard FLAME model, along with other Gaussian attributes including color, opacity, rotation, and scale. By modeling rigid head regions such as the forehead and scalp---which are largely unaffected by expressions---the Static Branch achieves robust generalization to novel viewpoints and accurate identity preservation. These expression-invariant regions allow their Gaussian parameters to be precomputed once during preprocessing, thereby significantly improving real-time performance. 
In contrast, the Dynamic Branch specializes in deformable facial regions such as the mouth, eyes, and cheeks. It leverages a separate decoder that combines the lightweight dynamic identity code $\mathbf{z}_c$ with the expression latent vector $\mathbf{z}$ from the displacement VAE to regress expression-dependent Gaussian parameters. This division of labor allows each branch to focus on its specialized role, avoiding the fidelity loss commonly observed in previous approaches that attempt to model both identity and expression within a monolithic framework. Finally, the outputs of the two branches are seamlessly blended, ensuring smooth transitions between static and dynamic regions.

% \noindent\textbf{Integration of 2D Vision Priors with 3D Data.}  
Beyond disentanglement, a second key insight of SEGA lies in effectively bridging the gap between 2D identity diversity and 3D geometric consistency. For static head regions, we adopt a DINOv2 backbone pretrained on large-scale 2D image collections to extract robust, general-purpose identity features. These features are further aligned into the UV space using a large reconstruction model, enabling the prediction of continuous, UV-aware Gaussian attributes that preserve fine-grained identity cues. For dynamic facial regions, we utilize a VQ-VAE encoder pretrained on large-scale 2D face datasets to produce discrete, codebook-quantized identity codes, which provide strong generalization across diverse identities. Meanwhile, 3D consistency is reinforced through joint training with multi-view, multi-expression 3D datasets, and further refined by a displacement VAE that predicts per-vertex geometric offsets beyond the standard FLAME topology. This integration of complementary 2D and 3D priors ensures that SEGA benefits from both the rich identity variation of 2D datasets and the geometric accuracy of 3D data.  
Finally, to enhance person-specific fidelity, SEGA performs a one-time fine-tuning on the input image, after which the photorealistic avatar can be rendered from arbitrary viewpoints using Gaussian Splatting.

In summary, our contributions include:
\begin{itemize}
    % overall contri. + experiments
    \item We propose SEGA, a novel method for high-quality, fully 360-degree renderable 3D avatar creation from a single image. Experiments demonstrate that SEGA outperforms existing methods in generalization, visual fidelity, and computational efficiency. 
    \item We design a hierarchical static–dynamic decomposition. The static branch handles rigid regions for identity preservation and novel-view generalization, while the dynamic branch models deformable regions for high-fidelity, real-time expression animation. 
    % This disentanglement avoids the fidelity loss of monolithic approaches.
    % contri.2: dual 2D priors + 3D data integration
    \item We fuse large-scale 2D priors (DINOv2, CodeFormer encoder) with multi-view 3D supervision and displacement VAE refinement, ensuring both rich identity diversity and geometric consistency. This integration enables SEGA to generalize across identities, viewpoints, and expressions.
    % % contri.3: hierarchical static-dynamic decomposition
    % \item We design a hierarchical UV-space Gaussian Splatting framework with static-dynamic decomposition that enables pre-computation of expression-independent regions for real-time performance while capturing fine-grained expression-dependent deformations. This decomposition maximizes data efficiency on scarce 3D datasets and ensures seamless facial animation capabilities.
\end{itemize}

\section{Related Work}
\label{sec:related}

%
%%%%%%%%%%%%%%%%%%%%%
%
%%%%%%%%%%%%%%%%%%%%%
%
\begin{figure*}[ht]
    \centering
    \includegraphics[width=\textwidth]{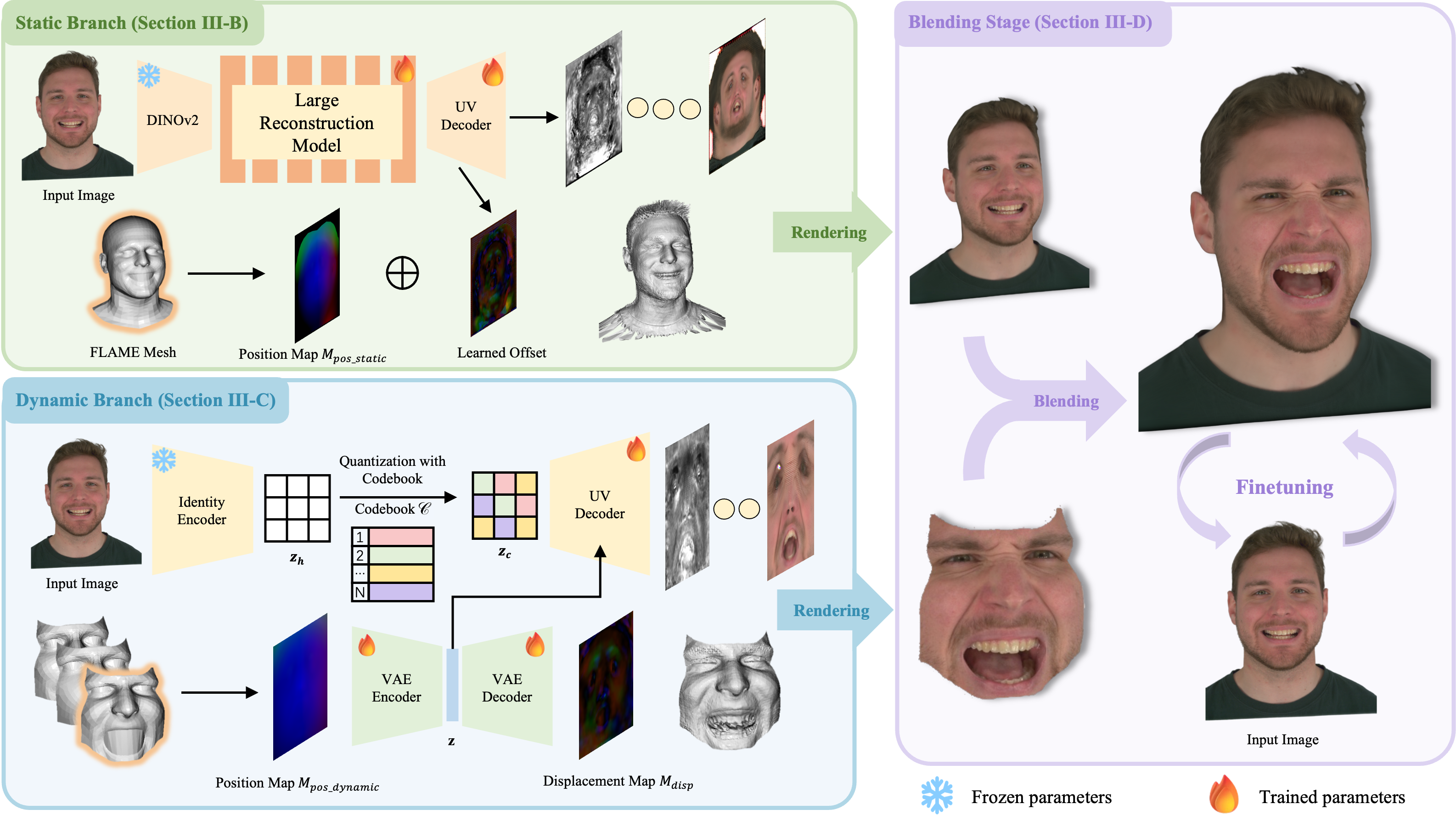}
    \caption{
    \textbf{Overview of SEGA method.}
    % 我们的方法包含3个部分：静态分支 动态分支 和 最后个性化stage
    % 在静态分支中，我们使用一个large pretrained image encoder （Dinov2）来提取信息。然后，将这些信息通过large reconstruction model （LRM）来fuse到uv space上。最后，我们用uv decoder解码出各个3dgs的属性。其中包括learned offset，可以apply在flame上，得到富有细节的几何和其他属性（scale ，opacity，color，rotation）
    % 在动态分支中， 我们使用一个pretrained VQ-VAE encoder to generate discrete identity code $\mathbf{z}_c$ for expression-dependent facial regions， and employs a VAE architecture to disentangle expression and identity information并且predict出在flame几何基础上的displacement map。最后动态分支的uv decoder combines expression latent vector $\mathbf{z}$ from the displacement VAE with dynamic identity code $\mathbf{z}_c$ 来decode出3dgs的各个属性。
    %最后，我们有一个个性化的分支。我们将运动的face region和静态部分blend成一个整体，然后使用输入的单id图片对整体finetune，得到最后结果。
    Our method consists of three parts: 
    \textbf{(1) Static Branch (\Cref{sec:static}):} We leverage a large pretrained image encoder (DINOv2) to extract identity features, which are fused into the UV space by a Large Reconstruction Model (LRM). A UV decoder then reconstructs various 3D Gaussian attributes, including a learned offset that can be applied to the FLAME mesh to enhance geometric details and other properties (scale, opacity, color, rotation). 
    \textbf{(2) Dynamic Branch (\Cref{sec:dynamic}):} We employ a pretrained VQ-VAE to obtain a discrete identity code $\mathbf{z}_c$ for expression-dependent regions, and use a VAE-based encoder--decoder to disentangle identity and expression, predicting displacement maps on top of the FLAME geometry. Then the UV decoder integrates the expression latent vector $\mathbf{z}$ with the identity code $\mathbf{z}_c$ to produce dynamic attributes of the 3D Gaussians. 
    \textbf{(3) Blending Stage (\Cref{sec:person}):} The static and dynamic components are blended into a unified representation, which is further finetuned with the input ID image to yield personalized, high-fidelity results.
    \vspace{-1em}
    }
    \label{fig:pipeline}
\end{figure*}
%
%%%%%%%%%%%%%%%%%%%%%
%
\noindent\textbf{Photorealistic 3D Avatar.}
Recently, the creation of animatable photorealistic 3D avatars has gained significant attention due to its potential applications in VR/AR and digital entertainment. 
Some methods employ mesh-based approaches to reconstruct avatars from individuals captured with multi-view RGB cameras~\cite{lombardi2018deep, UnstructureLan, su2022robustfusionPlus, robustfusion, jiang2022neuralhofusion}.
Some methods are driven by the recent development of neural implicit representations. 
Some leverage template mesh-guided canonical space NeRF framework~\cite{athar2022rignerf, gao2022reconstructing, xu2023avatarmav, zhao2023havatar, zheng2022imavatar, zielonka2023instant, jiang2023instant, zheng2024ohta}. Techniques such as INSTA~\cite{zielonka2022towards}, IM Avatar~\cite{zheng2022imavatar}, HAvatar~\cite{zhao2023havatar}, and PointAvatars~\cite{zheng2023pointavatar} deform points beyond template constraints via nearest neighbor strategies, whereas our method deforms the mesh template directly. 
Some other methods use the template-free approach. LatentAvatar~\cite{xu2023latentavatar} enhances expression transfer with latent codes, while Nersemble~\cite{kirschstein2023nersemble} reconstructs dynamic heads with multi-resolution hash grids, albeit lacking controllability.
Some methods like MVP~\cite{lombardi2021mixture}, and TRAvatar~\cite{yang2023towards} introduce volumetric primitive models for real-time rendering while relying on multi-view setups.

Recently, point-based representations such as 3D Gaussian Splatting (3DGS)~\cite{kerbl20233d} have gained attention due to their efficiency and high quality in modeling dynamic scenes~\cite{jiang2024hifi4g, jiang2025reperformerimmersivehumancentricvolumetric}. 
Recent methods regress the Gaussian parameters on the mesh surface~\cite{ma20243d, qian2024gaussianavatars, rivero2024rig3dgs, xu2024gaussian} and UV space~\cite{lan2023gaussian3diff, pang2024ash, saito2024rgca}, allows for dynamic control via latent codes~\cite{saito2024rgca} or expression parameters~\cite{qian2024gaussianavatars}, and even text~\cite{zhou2024headstudio}. 
Recent advances have further pushed the boundaries: GeoAvatar~\cite{moon2025geoavataradaptivegeometricalgaussian} presents adaptive geometrical Gaussian Splatting with automatic rigid-flexible segmentation, and FATE~\cite{zhang2025fatefullheadgaussianavatar} achieves the first animatable 360° full-head reconstruction from monocular video with neural baking techniques. DAGSM~\cite{zhuang2025dagsmdisentangledavatargeneration} combines mesh with 2D Gaussians for semantic separation, while LAM~\cite{he2025LAM} introduces large-scale model architectures for single-image animatable head generation. Privacy-preserving methods~\cite{wilson2025privacypreservingphotorealisticselfavatarsmixed} address de-identification for mixed reality applications.
\textcolor{revcolor}{UniGAHA~\cite{teotia2025unigaha} presents an audio-driven universal Gaussian head avatar framework, but requires per-subject video optimization.}
Although many of these works focus on the generation of single-subject avatars from multi-view inputs~\cite{qian2024gaussianavatars, saito2024rgca, xu2024gaussian}, our approach proposes a generalizable method for creating photo-realistic 3D avatars from a single image.

\noindent\textbf{One-shot 2D Head Avatar Synthesis.}
Recent years have seen growing interest in creating 2D head avatars, particularly for synthesizing realistic talking heads. While some researchers have developed methods using 2D generative models to animate static images through learned latent deformations~\cite{wiles2018x2face, siarohin2019first, ren2021pirenderer,pham2024style,wang2024styletalk++, ji2024realtalk}, these approaches often fail to maintain geometric consistency when handling different head poses and expressions in 3D space. Some studies have attempted to address this by incorporating NeRF-based methods to add 3D awareness~\cite{zhang2023metaportrait, peng2024synctalk, ye2024real3d}, though their results are still predominantly 2D in nature. 
The field has undergone a paradigm shift with the adoption of diffusion transformers. HunyuanPortrait~\cite{xu2025hunyuanportrait} establishes a new standard with its Stable Video Diffusion backbone and attention-based adapters for superior temporal consistency. Hallo3~\cite{cui2024hallo3} represents the first transformer-based video generative model for portrait animation with speech audio conditioning. FantasyPortrait~\cite{wang2025fantasyportrait} breaks new ground with multi-character support through Expression-augmented Diffusion Transformers and masked cross-attention mechanisms.
Although these techniques can produce realistic images, they lack true three-dimensional consistency, which makes them less suitable for applications in VR/AR setups.

\noindent\textbf{Few-shot 3D Head Avatar Creation.}
Creating personalized 3D head avatars from limited data has become a key research focus, driven by large-scale 3D datasets. Recent methods combine 3D priors with generative techniques for photorealistic avatars from minimal inputs. For instance, Morphable Diffusion~\cite{chen2024morphable} integrates diffusion models with multi-view consistency, while Preface~\cite{buhler2023preface} uses a NeRF-based approach for high-resolution avatars from sparse views, though it lacks animation support.
PhoneScan~\cite{cao2022authentic} generates a high-fidelity animatable 3D avatar using a phone capture of the entire head. However, the need for a dedicated phone scan limits practicality, and latent vector-based expression encoding restricts control over the head pose and expressions.
Recently, more and more papers use the 3D morphable model (3DMM)-based approaches like One2Avatar~\cite{yu2024one2avatar}, VRMM~\cite{yang2024vrmm}, Portrait4D~\cite{deng2024portrait4d}, Portrait4DV2~\cite{deng2024portrait4dv2}, Real3D~\cite{ye2024real3d}, GPAvatar~\cite{chu2024gpavatar}, GAGAvatar~\cite{chu2024generalizable} and HeadGAP~\cite{zheng2024headgap}, which have advanced the creation of photo-realistic avatars from a few input images or even one single image. 
The latest advances  have introduced generalizable priors: Avat3r~\cite{kirschstein2025avat3rlargeanimatablegaussian} combines Vision Transformer backbone with DUSt3R and Sapiens foundation models for high-quality reconstruction from just 4 temporal frames, and FaceLift~\cite{lyu2025faceliftlearninggeneralizablesingle} achieves 360-degree reconstruction from a single image through multi-view generation and GS-LRM. Additionally, synthetic prior approaches~\cite{zielonka2025syntheticpriorfewshotdrivable} have demonstrated the effectiveness of leveraging synthetic training data for few-shot drivable head avatar inversion, showing promising results in combining synthetic and real data for improved generalization. These methods leverage foundation model integration to significantly reduce training data requirements.
\textcolor{revcolor}{Beyond static or few-shot reconstruction, recent works have also explored 4D portrait avatar generation. CAP4D~\cite{taubner2025cap4d} creates animatable 4D portrait avatars using morphable multi-view diffusion models, and FaceCraft4D~\cite{yin2025facecraft4d} generates animated 3D facial avatars from a single image by leveraging temporal consistency in the 4D domain.}
However, these methods struggle to generalize across novel viewpoints, expressions, and identities, limiting their scalability and practical applications.

To overcome these limitations, our method introduces a novel end-to-end framework that leverages both 2D and 3D priors to achieve high-fidelity, drivable 3D head avatars from a single image. By disentangling identity and expression features, our approach ensures improved generalization to unseen subjects and expressions while maintaining multi-view consistency.

\section{Method}

In this section, we introduce SEGA, a novel framework for generating drivable 3D Gaussian avatars from a single input image.
The overall pipeline is illustrated in \Cref{fig:pipeline}.

The core idea of SEGA is two-fold. 
First, it disentangles identity and expression information through a hierarchical static–dynamic decomposition, which allows pre-computation of static components while ensuring real-time performance for dynamic facial animation. 
Second, it leverages 2D identity-consistency priors together with 3D geometry-consistency priors to generate high-quality 3D head avatars.

% The core idea of SEGA 一个是 to disentangle identity and expression information through a hierarchical static-dynamic decomposition. This hierarchical decomposition enables pre-computation of static components while maintaining real-time performance for dynamic facial animation. 
% 另一点是我们结合2D数据的ID一致性先验和3D数据的几何一致性先验来to generate high-quality 3D head avatar。
% %We leverage complementary 2D and 3D priors to achieve high-quality avatar generation while maintaining real-time performance during animation. 

Given a single human face image $I$, SEGA operates through three key parts. 
% A static branch to generate a full head 3dgs assets to 还原ID信息
% A dynamic branch to 实时更新动态的表情区域并维持高保真度。
%一个personalization stage，融合前两个branch的信息，并一步提升结果。
First, a \textbf{Static Branch (\Cref{sec:static})} generates full-head 3D Gaussian assets to faithfully preserve identity information.
Second, a \textbf{Dynamic Branch (\Cref{sec:dynamic})} updates expression-related facial regions in real time while maintaining high fidelity. 
Finally, a \textbf{Blending Stage (\Cref{sec:person})} fuses the outputs of the two branches and further refines the results using the input image, leading to high-quality and personalized 3D avatars.

\subsection{Preliminary} \label{sec:preliminary}
% \subsubsection{3D Gaussian Splatting}

In this work, we build the 3D human face avatar with 3D Gaussian Splatting~\cite{kerbl20233d}, which presents the face model with 3D Gaussian primitives. Centered at $\mathbf{p}$ and with covariance matrix as $\Sigma$, the 3D Gaussian primitive can be represented as $g(x) = \exp(-\frac{1}{2}(\mathbf{x} - \mathbf{p})^T\Sigma^{-1}(\mathbf{x} - \mathbf{p}))$. 
The covariance matrix $\Sigma$ is parametrized by the rotation quaternion $\mathbf{r}$ and scale vector $\mathbf{s}$ with $\Sigma = RSS^{-1}R^{-1}$, where $R$ and $S$ are the matrix representation of $\mathbf{r}$ and $\mathbf{s}$ respectively.

Next, the 3D Gaussian primitives are projected on the 2D camera space with EWA splatting~\cite{zwicker2002ewa}, resulting in the 2D Gaussians with the covariance matrix $\Sigma' = JW\Sigma W^TJ^T$, where the $W$ is the transformation matrix from the global coordinate system to the camera coordinate system, and $J$ is the Jacobian matrix of the camera projection function. 
Finally, the human face can be rendered from the splatted Gaussians following the point-based rendering:
% Finally, the splatted Gaussians are rendered into color $\mathbf{C}$ following the point-based rendering:
\begin{equation}
    \label{eq:gaussian_render}
    \mathbf{C}_\text{img} = \sum_{i\in N}\mathbf{c_i}a_i \prod_{j=i}^{i-1} (1-a_j).
\end{equation}
where $\mathbf{C}_\text{img}$ is the image color, $N$ is the number of Gaussians, $\mathbf{c_i}$ is the color of each Gaussian and $a_i$ is obtained by multiplying the covariance matrix $\Sigma'$ with the Gaussian opacity $o_i$.

\subsection{Static Branch} \label{sec:static}
%为了获得一个高精度的3D全头avatar并且有一个很好的id还原性，我们使用了一个3D的large reconstruction model 巧妙的将2D prior（2D pretrained image encoder）和3D数据融合。

Capturing fine-grained facial details while maintaining generalization across diverse human identities is challenging. Previous approaches often struggle because they are either trained on datasets with limited identity diversity~\cite{chu2024gpavatar, zheng2024headgap}, or their encoders lack sufficient capacity to discriminate identity-specific features. To address this issue, we integrate complementary 2D and 3D priors by adopting a DINOv2 backbone~\cite{oquab2023dinov2} as the 2D prior together with a large 3D reconstruction model (LRM). DINOv2, pretrained on large-scale 2D image collections, provides robust general-purpose representations, while the LRM possesses the capacity to effectively fuse these 2D features into our target UV space.  

\vspace{1mm}
\noindent\textbf{Network Architecture.~}
Given an input image $I$, DINOv2 encodes it into patch-wise feature tokens $\mathbf{F}$. To map these identity-aware features to the UV space, we employ a UV-Alignment Transformer~\cite{idol}, which serves as a large reconstruction model to fuse the image features $\mathbf{F}$ into our target UV tokens $\mathbf{z}_{s0}$. Specifically, $\mathbf{F}$ and $\mathbf{z}_{s0}$ are concatenated and passed through 10 transformer layers. Each layer applies unimodal self-attention, enabling effective information exchange between the two token sets. The final layer outputs the UV tokens, which serve as our static identity embedding $\mathbf{z}_s$.

A unified static UV Gaussian decoder $D_\text{static}$ then predicts all static Gaussian attributes in the UV space. Formally, given $\mathbf{z}_s$, the decoder outputs RGB color $\mathbf{c}_s \in \mathbb{R}^{H \times W \times 3}$, opacity $\mathbf{o}_s \in \mathbb{R}^{H \times W \times 1}$, rotation quaternion $\mathbf{r}_s \in \mathbb{R}^{H \times W \times 4}$, scale $\mathbf{s}_s \in \mathbb{R}^{H \times W \times 3}$, and position offset $M_{\text {offset}}(u, v) \in \mathbb{R}^{H \times W \times 3}$, where $H = W = 1024$ denotes the full-resolution size of the FLAME UV map:
\begin{equation}
    \text{GS}_{\text{static}} = D_{\text{static}}(\mathbf{z}_s).
\end{equation}
The opacity $\mathbf{o}_s$ is activated by a sigmoid to constrain values within $[0, 1]$, and the rotation quaternion $\mathbf{r}_s$ is normalized to ensure valid rotations:
\begin{equation}
    \mathbf{o}_s = \sigma(\tilde{\mathbf{o}}_s), \quad \mathbf{r}_s = \frac{\tilde{\mathbf{r}}_s}{\lVert \tilde{\mathbf{r}}_s \rVert_2}.
\end{equation}

These static Gaussian parameters are expression-independent and are designed to represent rigid head regions that remain invariant across expressions. Importantly, they can be \textbf{pre-computed once during preprocessing} and reused across all animation frames, except for the deformable facial regions.
We train our static branch on both 2D dataset FFHQ \cite{ffhq} and 3D multiview dataset.

\vspace{1mm}
\noindent\textbf{Offset Regression on the Standard FLAME Model.~}
Since the 3D Gaussian blobs are bound to the surface of the 3D human head mesh, obtaining high-quality human head geometry is crucial for achieving photorealistic rendering from the 3D Gaussians. To capture identity-specific geometric details beyond the standard FLAME model, we predict a static offset map $M_{\text{offset}}$ in UV space through our static UV Gaussian decoder $D_\text{s}$.

Specifically, the offset map is defined as $M_{\text{offset}}(u, v) = (\Delta x_s, \Delta y_s, \Delta z_s)$, where $\Delta x_s$, $\Delta y_s$, and $\Delta z_s$ represent identity-specific displacements of the 3D positions on the FLAME mesh surface. These offsets are directly decoded from the static identity embedding $\mathbf{z}_s$ as part of the static Gaussian parameters:
\begin{equation}
    [\mathbf{c}_s, \mathbf{o}_s, \mathbf{r}_s, \mathbf{s}_s, M_{\text{offset}}] = D_\text{static}(\mathbf{z}_s),
\end{equation}
where $M_{\text{offset}} \in \mathbb{R}^{H \times W \times 3}$ with $H = W = 1024$ representing the full-resolution UV map.

The static offsets capture person-specific geometric variations in expression-invariant regions (e.g., hair, facial contours, neck), enhancing the identity-specific details that remain constant across different expressions. These offsets are applied to the corresponding static regions of the static position map $M_\text{pos\_static}$ (derived from FLAME parameters). Unlike the dynamic displacement $M_\text{disp}$ which handles expression-dependent facial regions, $M_{\text{offset}}$ focuses exclusively on improving geometric fidelity in static head regions. We modify the FLAME topology and transform it into UV space to learn these geometry deltas. Detailed information about the modified FLAME mesh topology is provided in the supplementary material.

% % motivation 
% Capturing detailed
% facial features and ensuring generalization across diverse human identi-
% ties, 这是因为 they either train on datasets with limited identity diversity~\cite{chu2024gpavatar, zheng2024headgap} or 他们的encoder没有足够的能力分辨理解这些id。 To address this limitation, we 融合2D prior 和3D数据 by adopting a DINOv2 \cite{dinov2} backbone （2D prior） with a 3D large reconstruction model (LRM) . DINOv2, pretrained on large-scale 2D data, provides robust general visual representations, while the LRM 有充足的能力将这些2D信息fuse到我们的目标uv space上。

% %model
% Given an input image $I$, DINOv2 will encode it into patch-wise feature tokens, as $\textbf{F}$. Then to map this id image information to our target UV space, we employ a UV-Alignment Transformer \cite{idol} , a kind of large reconstruction model, to fuse the image feature $\textbf{F}$ into our target uv token $\mathbf{z}_{s0}$. Specifically, $\textbf{F}$ and $\mathbf{z}_{s0}$ are concat together to feed through 10 transformer layers. Each transformer layer is a 单模态 self attention to enable 信息间的相互传递。最后一层layer输出uv token 将会作为我们的static identity embedding $\mathbf{z}_s$。

%offset & deformation

% \input{figures/self_reenactment/Self}

\subsection{Dynamic Branch} \label{sec:dynamic}

To enable real-time expression-driven animation while preserving high-fidelity appearance, we leverage a lightweight VQ-VAE encoder~\cite{zhou2022towards}, pretrained on large-scale 2D face datasets~\cite{karras2019style}, as the dynamic identity encoder $E_\text{code}$ to extract face-specific features.  

Formally, the encoder $E_\text{code}$ maps an input image $I$ into a compressed feature map $\mathbf{z}_h \in \mathbb{R}^{m \times n \times d}$, which is subsequently vector-quantized. For each feature $\mathbf{z}_h^{i, j} \in \mathbb{R}^d$ at location $(i, j)$, where $i = 0, \ldots, m$ and $j = 0, \ldots, n$, we select the closest entry $\mathbf{z}_c^{i, j} \in \mathbb{R}^d$ from a pretrained codebook $\mathcal{C} = \left\{\mathbf{c}_k \in \mathbb{R}^d \mid k = 0, 1, \ldots, N_\text{code} \right\}$ via:
\begin{equation}
    \mathbf{z}_c^{i, j} = \underset{\mathbf{c}_k \in \mathcal{C}}{\arg\min} \left\|\mathbf{z}_h^{i, j} - \mathbf{c}_k \right\|_2,
\end{equation}
where $N_\text{code}$ is the codebook size. The quantized feature map $\mathbf{z}_c$ is obtained by aggregating $\mathbf{z}_c^{i, j}$ across all spatial locations. The parameters of $E_\text{s}$, $E_\text{code}$, and the codebook $\mathcal{C}$ remain frozen during the training of SEGA.  

\vspace{1mm}
\noindent\textbf{Deformation Regression.}  
For dynamic facial regions, we learn expression-dependent deformations to capture subtle geometric variations, particularly around the eyes and mouth. We adopt a VAE network $D_\text{disp}$ to predict a dynamic displacement map in the UV space. The displacement map is defined as $M_\text{disp}(u, v) = (\Delta x_d, \Delta y_d, \Delta z_d)$, where $\Delta x_d$, $\Delta y_d$, and $\Delta z_d$ represent expression-driven offsets of the 3D surface positions of the FLAME mesh. The VAE encoder maps an expression position map $M_\text{pos\_dynamic}$ into a latent vector $\mathbf{z}$, and the decoder reconstructs the displacement map $M_\text{disp}$ from this latent representation.  

\vspace{1mm}
\noindent\textbf{UV Decoding.}  
After obtaining the dynamic identity code $\mathbf{z}_c$ and the expression latent vector $\mathbf{z}$, we regress expression-dependent Gaussian parameters in UV space.  

Directly regressing Gaussian parameters from the static identity embedding is insufficient, as it cannot capture variations induced by expressions. This limitation degrades rendering quality even when mesh deformations are considered. To overcome this issue, we introduce a dynamic Gaussian decoder $D_\text{dynamic}$, which takes both $\mathbf{z}_c$ (from the VQ-VAE) and $\mathbf{z}$ (from the displacement VAE) as inputs to regress dynamic Gaussian attributes. The network outputs opacity $\mathbf{o}_d \in \mathbb{R}^{h \times w \times 1}$, color $\mathbf{c}_d \in \mathbb{R}^{h \times w \times 3}$, rotation quaternion $\mathbf{r}_d \in \mathbb{R}^{h \times w \times 4}$, and scale $\mathbf{s}_d \in \mathbb{R}^{h \times w \times 3}$ for facial regions affected by expressions:
\begin{equation}
    \text{GS}_{\text{dynamic}} = D_{\text{dynamic}}(\mathbf{z}_c, \mathbf{z}).
\end{equation}
The opacity is constrained with a sigmoid activation, and the quaternion is normalized:
\begin{equation}
    \mathbf{o}_d = \sigma(\tilde{\mathbf{o}}_d), \quad \mathbf{r}_d = \frac{\tilde{\mathbf{r}}_d}{\lVert \tilde{\mathbf{r}}_d \rVert_2}.
\end{equation}
Here, $h = w = 400$ denote the UV resolution of the facial region. During animation, these parameters are computed in real time, enabling accurate capture and rendering of subtle expression dynamics in the generated avatars.

\subsection{Blending Stage} \label{sec:person}
To simultaneously preserve the fine details of a full-head avatar and support real-time animation, we seamlessly combine the static and dynamic branches. During inference, the static branch components are computed only once per identity and cached, while the dynamic branch is evaluated for each new expression, significantly reducing computational overhead.  

To obtain the final deformed mesh, we use the static position with offsets ($M_\text{pos\_static} + M_\text{offset}$) as the base, and replace the central facial region (400×400 UV area) with the dynamic deformation ($M_\text{pos\_static} + M_\text{disp}$). Using the pre-defined binary mask $M_\text{face}$, the final position map is computed as: $\hat{M}_\text{pos} = M_\text{face} \odot (M_\text{pos\_static} + M_\text{disp}) + (1-M_\text{face}) \odot (M_\text{pos\_static} + M_\text{offset})$. The final deformed mesh geometry $\mathbf{G}$ is then derived by sampling the modified position map $\hat{M}_\text{pos}$.

The final Gaussian parameters in the UV space are obtained by combining expression-dependent parameters in the facial region with expression-independent parameters in the remaining head regions. Using a pre-defined binary mask $M_\text{face}$ of the dynamic region of the face, we compute the final parameters of opacity $\mathbf{o}$, rotation quaternion $\mathbf{r}$, and scale $\mathbf{s}$ as:
% $\mathbf{o} = M_\text{face} \odot \mathbf{o}_d + (1-M_\text{face}) \odot \mathbf{o}_s$; $\mathbf{r} = M_\text{face} \odot \mathbf{r}_d + (1-M_\text{face}) \odot \mathbf{r}_s$; $\mathbf{s} = M_\text{face} \odot \mathbf{s}_d + (1-M_\text{face}) \odot \mathbf{s}_s$,
\begin{equation}
\begin{split}
    \mathbf{o} &= M_\text{face} \odot \mathbf{o}_d + (1-M_\text{face}) \odot \mathbf{o}_s \\
    \mathbf{r} &= M_\text{face} \odot \mathbf{r}_d + (1-M_\text{face}) \odot \mathbf{r}_s \\
    \mathbf{s} &= M_\text{face} \odot \mathbf{s}_d + (1-M_\text{face}) \odot \mathbf{s}_s \\
\end{split}
\end{equation}
where the $\odot$ is the Hadamard product. 
% In order to ensure smooth transitions between the dynamic facial regions and the static head regions, we use Poisson blending~\cite{perez2023poisson} to obtain the color parameter $\mathbf{c}$ in the UV space:
% \begin{equation}
%     \mathbf{c} = \mathcal{F}_\text{Poisson}(M_\text{face}, \mathbf{c}_d, \mathbf{c}_s)
% \end{equation}
% where the $\mathcal{F}_\text{Poisson}$ is the Poisson blending function.
To ensure smooth transitions between the dynamic facial regions and the static head regions, we construct a transition area along the edge of the target region and generate a weight mask \(\mathbf{M}_\text{f} \) through linear interpolation. For the static color map \( \mathbf{c}_s \) and dynamic color map \( \mathbf{c}_d \), the fused color map \( \mathbf{c} \) can be expressed as: $\mathbf{c} = \mathbf{M}_\text{f}\mathbf{c}_d + (1-\mathbf{M}_\text{f})\mathbf{c}_s$, 
where \( \mathbf{M}_\text{f} \) smoothly transitions from 0 to 1 within the transition band, remains 1 inside the region of dynamic color map \( \mathbf{c}_d \), and stays 0 outside the region. This gradient-weight design ensures smooth fusion of the images in the transition area, effectively avoiding visible seams at the stitching boundaries.

\vspace{1mm}
\noindent{\textbf{Rendering Human Head Avatar.~}} 
To generate 3D Gaussian primitives from the precomputed UV-space attributes, we employ a structured sampling strategy on the UV map. Given the UV-space Gaussian parameters $\mathbf{A} \in \mathbb{R}^{H \times W \times D}$, where $H = W = 1024$ and $D = 11$ encodes the complete set of Gaussian attributes (color, opacity, rotation, scale, and position offset), we perform regular grid sampling with step size $s$ to extract discrete Gaussian primitives.

Specifically, we partition the UV space into square grids and extract two triangular primitives per grid. For each grid cell with corner coordinates $(i, j)$, $(i+s, j)$, $(i, j+s)$, and $(i+s, j+s)$, where $i, j \in \{0, s, 2s, ...\}$, we compute the Gaussian attributes through barycentric averaging:
\begin{equation}
\begin{split}
    \text{GS}_1^{(i,j)} &= \frac{1}{3}\left[\mathbf{A}(i,j) + \mathbf{A}(i+s,j) + \mathbf{A}(i,j+s)\right] \\
    \text{GS}_2^{(i,j)} &= \frac{1}{3}\left[\mathbf{A}(i+s,j) + \mathbf{A}(i,j+s) + \mathbf{A}(i+s,j+s)\right]
\end{split}
\end{equation}
where $\mathbf{A}(u,v)$ denotes the attribute vector at UV coordinate $(u,v)$, and $\text{GS}_k^{(i,j)}$ represents the $k$-th Gaussian primitive extracted from the grid cell at position $(i,j)$. This triangulation-based sampling ensures complete coverage of the UV space while maintaining computational efficiency.

The 3D position $\mathbf{p}_k$ of each Gaussian primitive is computed by averaging the corresponding 3D coordinates from the deformed mesh $\mathbf{G}$:
\begin{equation}
    \mathbf{p}_k = \frac{1}{3}\sum_{m=1}^{3} \mathbf{G}(u_m, v_m)
\end{equation}
where $(u_m, v_m)$ are the UV coordinates of the three vertices defining the triangular region. Following the standard 3D Gaussian Splatting pipeline (\Cref{sec:preliminary}), these primitives are projected onto the image plane and rasterized to produce the final rendered image $\hat{I}$ through alpha compositing.

\begin{figure}[t]
    \centering
    \includegraphics[width=\linewidth]{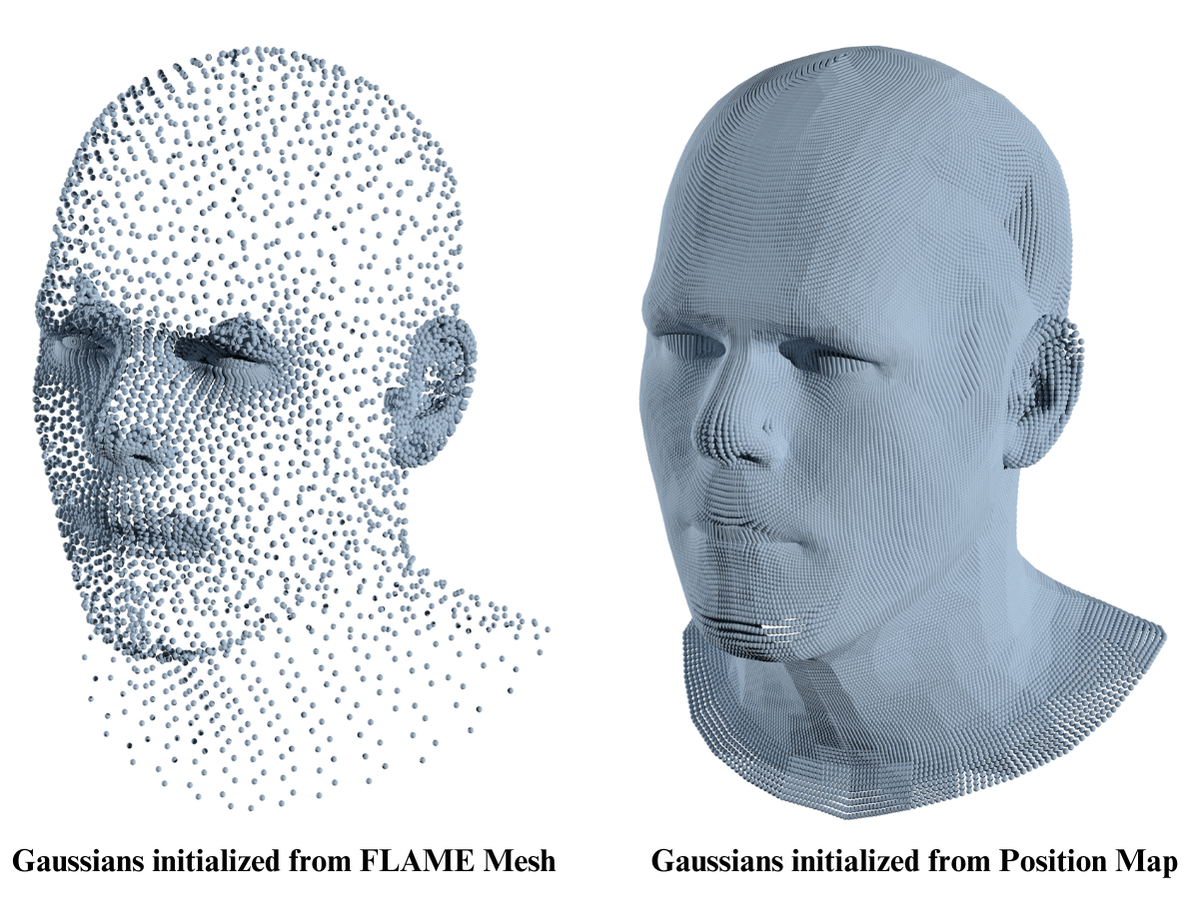}
    \caption{\textcolor{revcolor}{Gaussians initialized from FLAME mesh faces (left) versus from position map via regular UV grid sampling (right). Mesh-based initialization leads to non-uniform density, while UV grid sampling provides even coverage across the head.}}
    \label{fig:whypositionmap}
\end{figure}

\textcolor{revcolor}{Notably, we perform structured sampling on the regular UV grid rather than directly on the FLAME mesh triangles. The FLAME mesh topology is highly non-uniform---faces are densely concentrated around the eyes, nose, mouth, and ears, while the forehead, cheeks, and neck are covered by far fewer triangles. Initializing one Gaussian per mesh face therefore produces an overly dense distribution in the former regions and sparse coverage elsewhere, as illustrated in \Cref{fig:whypositionmap}, hindering both training convergence and detail reconstruction. By contrast, regular UV grid sampling yields a spatially uniform Gaussian distribution that covers the entire head surface evenly, leading to more balanced optimization and better fidelity. Furthermore, the UV-space representation naturally encodes all geometric and appearance information as three-channel attribute maps, which can be efficiently processed by standard CNN and transformer decoders.}

\begin{figure*}[htbp!]
    \centering
    \includegraphics[width=\textwidth]{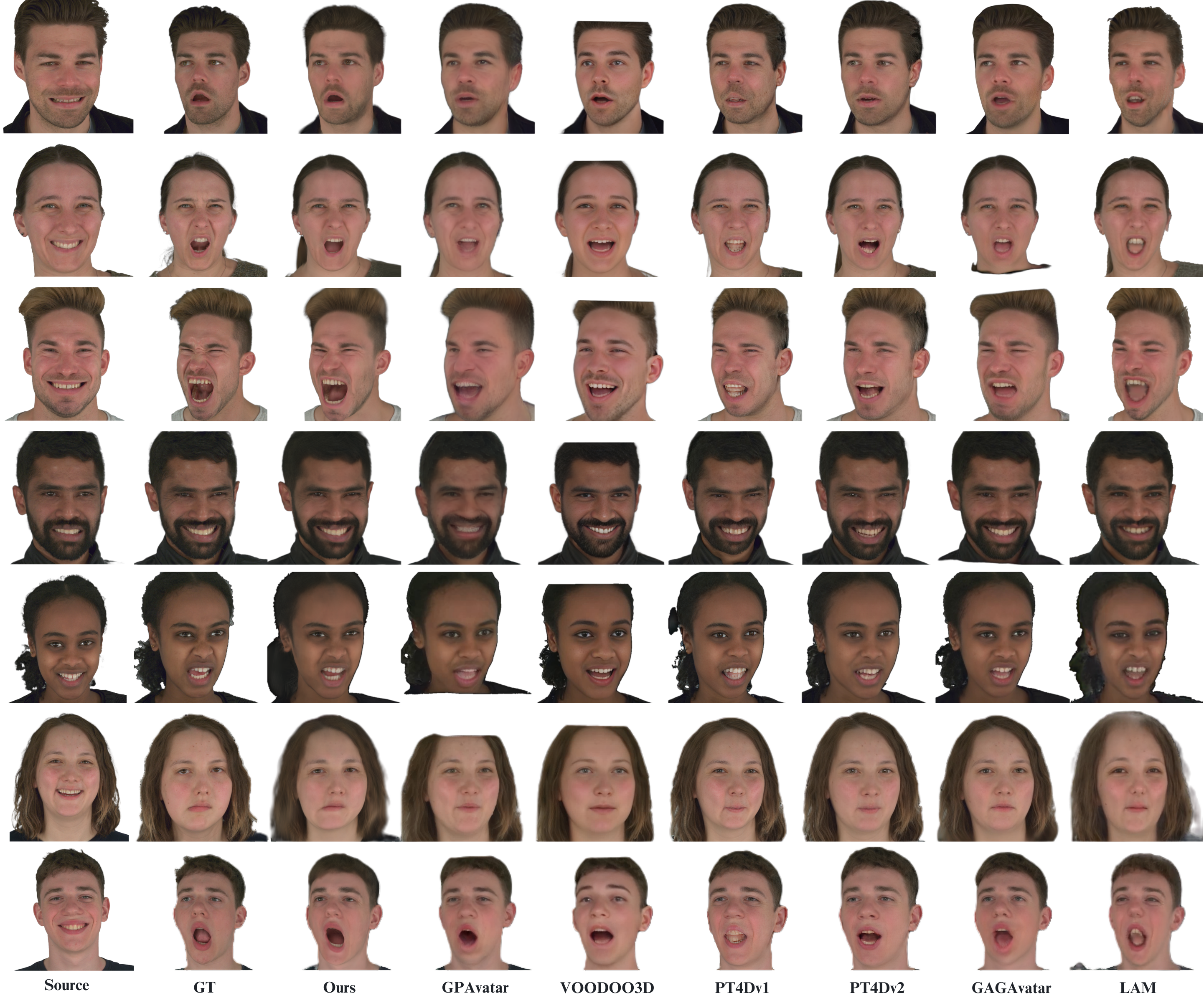}
    % \vspace{-3mm}
    \caption{Qualitative comparison 
    % of self-reenactment results 
    on NeRSemble dataset. SEGA demonstrates superior facial expression/pose fidelity and identity preservation compared to SOTA methods, including LAM~\cite{he2025LAM}, GPAvatar~\cite{chu2024gpavatar}, VOODOO3D~\cite{tran2023voodoo}, PT4Dv1~\cite{deng2024portrait4d}, PT4Dv2~\cite{deng2024portrait4dv2} and GAGAvatar~\cite{chu2024generalizable}.}
% \vspace{-5mm}
\label{fig:self-re}
\end{figure*}

\vspace{1mm}
\noindent{\textbf{Person-Specific Finetuning.~}} \label{sec:finetuning}
% To further enhance performance, 
To further capture fine-grained, identity-specific details, we perform a one-time person-specific fine-tuning on the input image.
During this process, we first extract the FLAME parameters from the input image $I$, and then fine-tune all trainable networks, including the static Gaussian decoder and the dynamic Gaussian decoder, using only MSE loss ($L_\text{L2}$) and perceptual loss ($L_\text{perc}$). 
This process is highly efficient and completed in just a few minutes. Notably, once fine-tuned, the avatar can be driven directly by FLAME parameters without further optimization.
% We note that person-specific fine-tuning needs to be performed only once for a new input image of a person. No additional fine-tuning is required when driving the avatar using FLAME parameters.

\subsection{Training Details}
% During the training process, we train the static Gaussian decoder $D_\text{static}$, dynamic Gaussian decoder $D_\text{dynamic}$, and displacement VAE network $D_\text{disp}$ simultaneously in an end-to-end manner. 
During the training process, we optimize the UV-Alignment Transformer, the static Gaussian decoder $D_\text{static}$ (full-resolution UV map, $H=W=1024$), the dynamic Gaussian decoder $D_\text{dynamic}$ (facial-region UV map, $h=w=400$), and the displacement VAE $D_\text{disp}$. The static and dynamic branches can be trained independently due to their modular design. For supervision, we adopt different data schedules per branch: for each identity, the static branch uses a curated multi-view set under a single teeth-exposing smile expression as ground truth, while the dynamic branch uses the full set expressions for that identity. Throughout training, the DINOv2 backbone, the dynamic identity encoder $E_\text{code}$, and the codebook $\mathcal{C}$ remain frozen.
For the training of the displacement VAE network $D_\text{disp}$, we designed the loss terms in \Cref{eq:train_vae}.
Following~\cite{kingma2013auto}, the training loss of the displacement VAE $D_\text{disp}$ is formulated as:
\begin{equation} \label{eq:train_vae}
\begin{split}
    L_\text{disp} &= \lambda_1 \left\| \hat{N} - N \right\|_2 + \lambda_2 L_\text{lap} (\mathbf{G}) \\
    &+ \lambda_3 L_\text{norm} (\mathbf{G}) + \lambda_4 KL\left(q(\mathbf{z} | M_\text{pos}) \| \mathcal{N}(0, I)\right)
\end{split}
\end{equation}
where $N$ is the ground truth normal map. We further introduce a Laplacian smoothness term $L_\text{lap} (\mathbf{G})$ \cite{desbrun1999implicit} and a normal consistency term $L_\text{norm} (\mathbf{G})$ \cite{ravi2020pytorch3d} to regularize the deformation. The normal consistency term aims to minimize the normal difference between two neighboring faces. $q(\mathbf{z} | M_\text{pos\_dynamic})$ refers to the projected distribution of input position map $M_\text{pos\_dynamic}$ in the latent space, $\mathcal{N}(0, I)$ refers to the standard normal distribution, and $KL(.)$ is the Kullback–Leibler divergence. The $\lambda_{1, ..., 4}$ are the weights of each loss term.

For the static branch training, we use only photometric losses to optimize the UV-Alignment Transformer and static Gaussian decoder $D_\text{static}$:
\begin{equation}
    L_\text{static} = \lambda_6 L_\text{L2} + \lambda_7 L_\text{perc}
\end{equation}

For the dynamic branch training, we combine both geometric losses (for the displacement VAE $D_\text{disp}$) and photometric losses (for the dynamic Gaussian decoder $D_\text{dynamic}$):

\begin{equation}
    L_\text{dynamic} = L_\text{disp} + \lambda_6 L_\text{L2}  + \lambda_7 L_\text{perc}
\end{equation}
where $L_\text{disp}$ is shown in \Cref{eq:train_vae}, and $\lambda_{6, 7}$ are the weights of the photometric losses. 
The $L_\text{L2}$ is the MSE loss between the rendered image $\hat{I}$ and the ground truth image $I_\text{gt}$ in the specific training camera: $L_\text{L2} = \left\| \hat{I} - I_\text{gt} \right\|_2$.
The $L_\text{perc}$ is the perceptual loss~\cite{johnson2016perceptual} between the rendered image $\hat{I}$ and ground truth image $I_\text{gt}$, leveraging the pretrained VGG network~\cite{simonyan2014very} to minimize the distance of VGG-extracted features between $\hat{I}$ and $I_\text{gt}$.

\section{Experimental}\label{sec:exp}

% \begin{table*}[t!]
%   \centering
%   \begin{tabularx}{\textwidth}{ll*{6}{X}}
%     \toprule
%     Method & Reference & PSNR$\uparrow$ & SSIM$\uparrow$ & LPIPS$\downarrow$ & CSIM$\uparrow$ & AKD$\downarrow$ & AED$\downarrow$ \\
%     \midrule
%     GPAvatar\cite{chu2024gpavatar} & ICLR 2024& 22.1675 & 0.7814 & 0.2722 & 0.8263 & 9.5921 & 3.8142 \\
%     VOODOO 3D\cite{tran2023voodoo} & CVPR 2024 & 22.4534 & 0.7659 & 0.2535 & 0.8351 & 9.1284 & 3.5967 \\
%     Portrait4D-v1\cite{deng2024portrait4d} & CVPR 2024 & 22.6196 & 0.7705 & 0.2434 & 0.8389 & 8.9267 & 3.9158 \\
%     Portrait4D-v2\cite{deng2024portrait4dv2} & ECCV 2024 & 22.7829 & 0.7727 & 0.2386 & 0.8435 & 8.7691 & 3.6783 \\
%     GAGAvatar\cite{chu2024generalizable} & NIPS 2024 & 22.4458 & 0.7967 & 0.2266 & 0.8474 & 8.6513 & 3.5924 \\
%     LAM\cite{he2025LAM} & CVPR 2025 & 22.9361 & 0.7883 & 0.2301 & 0.8459 & 8.5596 & 3.6132 \\
%     \rowcolor[rgb]{.906,.902,.902} Ours & - & \textbf{24.5862} & \textbf{0.8219} & \textbf{0.2236} & \textbf{0.8549} & \textbf{8.4127} & \textbf{3.3845} \\
%     \bottomrule
%     \end{tabularx}%
%     \caption{Quantitative Comparison on NeRSemble dataset\textcolor{revcolor}{, averaged over 6 camera viewpoints and 2 expression sequences}.}
%     \label{tab:main_results}%
% \end{table*}

\begin{table*}[t!]
  \centering
  \begin{tabularx}{\textwidth}{ll*{6}{X}}
    \toprule
    Method & Reference & PSNR$\uparrow$ & SSIM$\uparrow$ & LPIPS$\downarrow$ & CSIM$\uparrow$ & AKD$\downarrow$ & AED$\downarrow$ \\
    \midrule
    GPAvatar\cite{chu2024gpavatar} & ICLR 2024& 21.7549 & 0.7803 & 0.2945 & 0.7738 & 7.3509 & 3.5184 \\
    VOODOO 3D\cite{tran2023voodoo} & CVPR 2024 & 21.2027 & 0.7454 & 0.3009 & 0.7612 & 8.1092 & 3.5847 \\
    Portrait4D-v1\cite{deng2024portrait4d} & CVPR 2024 & 21.0742 & 0.6846 & 0.3068 & 0.7911 & 8.1801 & 4.3409 \\
    Portrait4D-v2\cite{deng2024portrait4dv2} & ECCV 2024 & 22.2162 & 0.7513 & 0.2675 & 0.8056 & 8.2347 & 3.7099 \\
    GAGAvatar\cite{chu2024generalizable} & NIPS 2024 & 23.1753 & 0.8022 & 0.2524 & 0.8345 & 6.2346 & 2.8229 \\
    LAM\cite{he2025LAM} & CVPR 2025 & 22.8130 & 0.7845 & 0.2917 & 0.8191 & 7.3451 & 3.0809 \\
    \rowcolor[rgb]{.906,.902,.902} Ours & - & \textbf{24.4944} & \textbf{0.8183} & \textbf{0.2519} & \textbf{0.8462} & \textbf{5.5751} & \textbf{2.8228} \\
    \bottomrule
    \end{tabularx}%
    \caption{Quantitative Comparison on NeRSemble dataset\textcolor{revcolor}{, averaged over 6 camera viewpoints and 2 expression sequences}.}
    \label{tab:main_results}%
\end{table*}
% \subsection{Experiment setup}

\subsection{Details}
\noindent\textbf{Dataset.}
We train our generalizable model using two comprehensive datasets: the NeRSemble~\cite{kirschstein2023nersemble} dataset and our captured dataset. For training, we utilize 110 subjects from NeRSemble and 1000 subjects from our captured dataset, while 34 subjects from NeRSemble are used for evaluation. Detailed information about dataset composition, preprocessing procedures, and FLAME tracking configurations is provided in the supplementary material.

\noindent\textbf{Training.}
The entire method is implemented using the PyTorch framework~\cite{paszke2019pytorch} and trained on 8 A100-80G GPUs. For rendering, we use gsplat~\cite{ye2024gsplatopensourcelibrarygaussian}. Training uses the Adam optimizer with a learning rate of $3 \times 10^{-4}$, and loss weights are set as $\lambda_1 = 1$, $\lambda_2 = 1$, $\lambda_3 = 0.1$, $\lambda_4 = 20$, $\lambda_5 = 0.01$, $\lambda_6 = 1000$, $\lambda_7 = 1000$. Training the static decoder (full-resolution: $H = 1024$, $W = 1024$) takes 36 hours, while the dynamic decoder (partial-resolution: $h = 400$, $w = 400$) requires 24 hours.

\noindent\textbf{Avatar Personalization.}
For avatar creation, we fine-tune the static and dynamic networks with loss weights set as  $\lambda_6 = 1000 $ and $\lambda_7 = 1000 $, which takes 2 minutes per branch on a single NVIDIA A100 GPU. 
As for avatar animation, it requires no additional training, and its runtime on a single GPU is \textbf{50ms} per frame, achieving real-time performance. 
% Specifically, the breakdown is as follows: dynamic GS generation (37.65ms), GS color fusion (0.34ms), position map sampling (0.23ms), and rendering (13.28ms).
Specifically, the detailed breakdown of the process is as follows: dynamic GS generation (37.65ms), GS color fusion (0.34ms), position map sampling (0.23ms), and rendering
(13.28ms).
% Experiments are conducted using the Nersemble dataset~\cite{kirschstein2023nersemble}, with 102 subjects for training and xx for testing. We also test on real-world data from 3 individuals captured using a smartphone. For FLAME registration and background removal, we follow HeadGAP~\cite{zheng2024headgap}, and align faces for single-image inputs as in CodeFormer~\cite{zhou2022towards}, using a white background for inpainting.

% \noindent\textbf{Dataset and Preprocessing}
% We conduct experiments using the Nersemble dataset~\cite{kirschstein2023nersemble} with xx camera viewpoints. We use 102 subjects for training, while xx subjects are reserved for testing. 
% Additionally, we test our method on in-the-wild data by capturing images of 3 individuals using a smartphone.
% For FLAME registration and background removal, we follow the approach from HeadGAP~\cite{zheng2024headgap}.
% For single image input, we align the input face images as CodeFormer~\cite{zhou2022towards}, using a white background to in-paint the cropped areas of the images.

\noindent\textcolor{revcolor}{\textbf{Data Preprocessing.}
For the NeRSemble dataset, we Background Matting V2~\cite{lin2021bgmv2} to perform foreground segmentation and set the background to black. For other data sources, we use MODNet~\cite{MODNet} for portrait matting, retaining only pixels with alpha values above 200 as foreground, which effectively suppresses semi-transparent boundary artifacts and yields cleaner subject edges. During evaluation, for fair comparison, we process the LAM~\cite{he2025LAM} outputs with MODNet using the same alpha thresholding to produce black-background images, while other baselines already output black backgrounds by default. All results are then aligned using face-alignment~\cite{bulat2017far} before metric computation, ensuring a unified evaluation protocol.}

\noindent\textcolor{revcolor}{\textbf{Evaluation Metrics.}
We adopt six metrics for quantitative evaluation: Peak Signal-to-Noise Ratio (PSNR) and Structural Similarity Index (SSIM)~\cite{wang2004image} for pixel-level and structural fidelity, Learned Perceptual Image Patch Similarity (LPIPS)~\cite{zhang2018unreasonable} for perceptual quality using VGG~\cite{simonyan2014very} features, Cosine Similarity (CSIM) using ArcFace~\cite{deng2019arcface} embeddings for identity preservation, Average Keypoint Distance (AKD) for facial landmark accuracy, and Average Expression Distance (AED) based on landmark centroid distance for expression transfer fidelity. Detailed formulations are provided in the supplementary material.}

\begin{figure*}[!ht]
    \centering
    \vspace{-2mm}
    \includegraphics[width=1\textwidth]{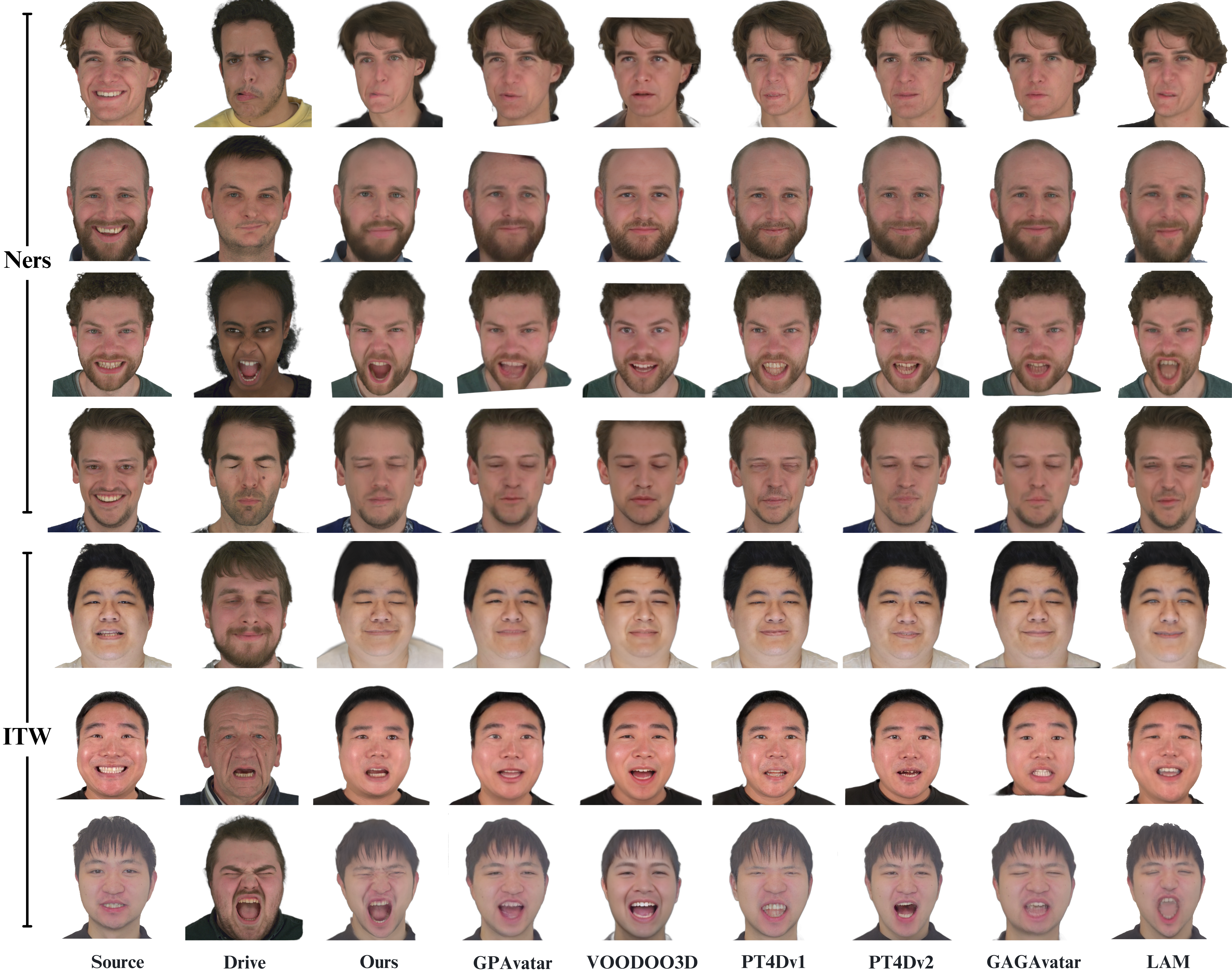}
    \caption{
    Cross-identity reenactment qualitative comparison across NeRSemble and in-the-wild data. We compare our method with state-of-the-art baseline approaches including GPAvatar, VOODOO3D, Portrait4D-v1, Portrait4D-v2, GAGAvatar, and LAM. The first four rows show results on NeRSemble dataset subjects with controlled studio conditions, while the bottom three rows demonstrate results on in-the-wild data captured using different smartphone models under varying lighting conditions. Each row shows cross-identity reenactment results where the source identity (leftmost column) is animated using driving expressions from different subjects. Our method consistently produces more accurate expression transfer while better preserving identity-specific features compared to baseline methods across both controlled and challenging real-world scenarios.
}
    \label{fig:cross}
    \vspace{-2mm}
\end{figure*}

\subsection{Comparison}
\noindent\textbf{Baselines.}
We compare our method with several state-of-the-art single-image head avatar generation approaches, including LAM~\cite{he2025LAM}, GPAvatar~\cite{chu2024gpavatar}, VOODOO3D~\cite{tran2023voodoo}, Portrait4D-v1~\cite{deng2024portrait4d}, Portrait4D-v2~\cite{deng2024portrait4dv2}, and GAGAvatar~\cite{chu2024generalizable}. Among them, LAM (Latent Avatar Modeling) leverages latent diffusion models with 3D-aware guidance to generate high-fidelity head avatars from single images, focusing on identity preservation and expression control through latent space manipulation. Portrait4D-v1 and Portrait4D-v2 adopt tri-plane-based representations for one-shot avatar synthesis. GPAvatar introduces a generalizable pipeline from single or multiple images, while VOODOO3D utilizes volumetric disentanglement and 3D frontalization for expression transfer. GAGAvatar proposes a dual-lifting strategy with 3DMM prior constraints to create generalizable and animatable 3D Gaussian head avatars from a single image.
\textcolor{revcolor}{All baselines are evaluated using their official pretrained models. Among them, LAM, GPAvatar, and GAGAvatar are trained on the VFHQ dataset, Portrait4D-v1 and Portrait4D-v2 are trained on FFHQ and VFHQ with synthetic multi-view augmentation, and VOODOO3D is trained on CelebV-HQ. Detailed training data descriptions for each baseline are provided in the supplementary material.}

\vspace{1mm}
\noindent\textbf{Self Reenactment Quantitative Comparison.}
For quantitative evaluation, we conduct comprehensive experiments on \textcolor{revcolor}{7} NeRSemble~\cite{kirschstein2023nersemble} subjects. \textcolor{revcolor}{All metrics are averaged over 6 camera viewpoints spanning frontal, near-frontal, and side views, and 2 expression sequences, ensuring comprehensive multi-view and multi-expression evaluation coverage. Detailed subject IDs, camera IDs, and expression sequences are provided in the supplementary material.} As presented in Table~\ref{tab:main_results}, our method demonstrates superior performance across all metrics, achieving the best results in PSNR (24.4944), SSIM (0.8183), LPIPS (0.2519), CSIM (0.8462), AKD (5.5751), and AED (2.8228), highlighting the effectiveness of our integrated 2D and 3D priors for comprehensive avatar quality and fidelity.

\vspace{1mm}
\noindent\textbf{Self Reenactment Qualitative Comparison.}  
As illustrated in Fig.~\ref{fig:self-re}, our method consistently achieves superior facial expression fidelity compared to other state-of-the-art approaches. Our results exhibit both precise expression matching and enhanced visual realism, corroborating the quantitative superiority indicated by all evaluation metrics in Table~\ref{tab:main_results}.

The visual comparisons in Fig.~\ref{fig:self-re} complement our quantitative analysis, showing that our method not only achieves superior numerical performance but also produces more realistic and expressive facial animations. The consistent quality across different subjects and expressions demonstrates the robustness of our hierarchical dynamic-static framework.

\vspace{1mm}
\noindent\textbf{Cross-Identity Reenactment Quantitative Comparison.}
To further validate the effectiveness of our method in cross-identity scenarios, we conduct comprehensive quantitative comparisons against state-of-the-art approaches on cross-identity reenactment tasks. \textcolor{revcolor}{We evaluate on 4 cross-identity pairs from NeRSemble, using the same 6 camera viewpoints and 2 expression sequences as in the self-reenactment evaluation (detailed pair IDs are provided in the supplementary material).} The cross-identity reenactment task presents additional challenges as it requires the model to disentangle identity-specific features from expression-dependent variations effectively.

% 强制在第5页显示quantitative表格
% \ifnum\value{page}<4
%   \afterpage{\afterpage{\afterpage{\afterpage{\input{Tabs/quantitative}}}}}
% \else
%   \input{Tabs/quantitative}
% \fi
% \begin{table*}[htbp]
%   \centering
%   \caption{Quantitative ablation study comparing the performance of different configurations.}
%   \begin{tabularx}{\textwidth}{l*{8}{X}}
%     % \begin{tabular}{ccccccccc}
%     \toprule
%     \multirow{2}[3]{*}{Method} & \multicolumn{3}{c}{Frontal view} & \multicolumn{3}{c}{All views} \\
% \cmidrule{2-7}          & PSNR  & SSIM  & \multicolumn{1}{l}{LPIPS}  & PSNR  & SSIM  & \multicolumn{1}{l}{LPIPS}  \\
%     \midrule
%     Full-static variation &       &       &       &        &       &  \\
%     Full-dynamic variation &       &       &       &          &       &  \\
    
%     \midrule
%     Baseline method  &       &       &       &       &         &  \\
%     - Identity loss  &       &       &       &       &        &  \\
%     \quad - Perceptual loss  &       &       &       &            &       &  \\
%     - 2D prior &       &       &       &       &         &  \\
%     \rowcolor[rgb]{ .906,  .902,  .902} + Finetuning (Full method)  &       &          &       &       &       &  \\
%     % w/o \cite{zheng2024headgap} &       &       &       &       &       &       &       &  \\
%     \bottomrule
    
%     % \end{tabular}%
%     \end{tabularx}%
%   \label{tab:ablation}%
% \end{table*}%

\begin{table}[!htbp]
  \centering

  \begin{tabularx}{0.45\textwidth}{l*{4}{X}}
    % \begin{tabular}{ccccccccc}
    \toprule
%     \multirow{2}[3]{*}{Method} & \multicolumn{3}{c}{Frontal view} & \multicolumn{3}{c}{All views} \\
% \cmidrule{2-7}          & PSNR  & SSIM  & \multicolumn{1}{l}{LPIPS}  & PSNR  & SSIM  & \multicolumn{1}{l}{LPIPS} 

    Method &    CSIM$\uparrow$  & AKD$\downarrow$  &AED$\downarrow$  \\
    \midrule
    GPAvatar\cite{chu2024gpavatar} &  0.8168     &  9.4417     & 3.5743     \\
    VOODOO3D\cite{tran2023voodoo} & 0.7345      & 8.2374      &  3.3534     \\
    Portrait4D-v1\cite{deng2024portrait4d} & 0.7809      &  8.0512    & 3.4284      \\
    Portrait4D-v2\cite{deng2024portrait4dv2}  &  0.8096     &  8.0246     &     3.3562 \\
    GAGAvatar\cite{chu2024generalizable} & 0.8438      &  7.9187           & 3.3464 \\
    LAM\cite{he2025LAM}  &  0.8265     &   8.1803        &  3.8561\\
    \rowcolor[rgb]{ .906,  .902,  .902} Ours  &  \textbf{0.8517}      &  \textbf{7.8183}     & \textbf{3.2697} \\
    % w/o \cite{zheng2024headgap} &       &       &       &       &       &       &       &  \\
    \bottomrule
    
    % \end{tabular}%
    \end{tabularx}%
      \caption{Quantitative Comparison on Cross-Identity Reenactment.}
  \label{tab:cross}%
\end{table}%

\begin{figure}[!htbp]
    \centering
    \includegraphics[width=0.95\linewidth]{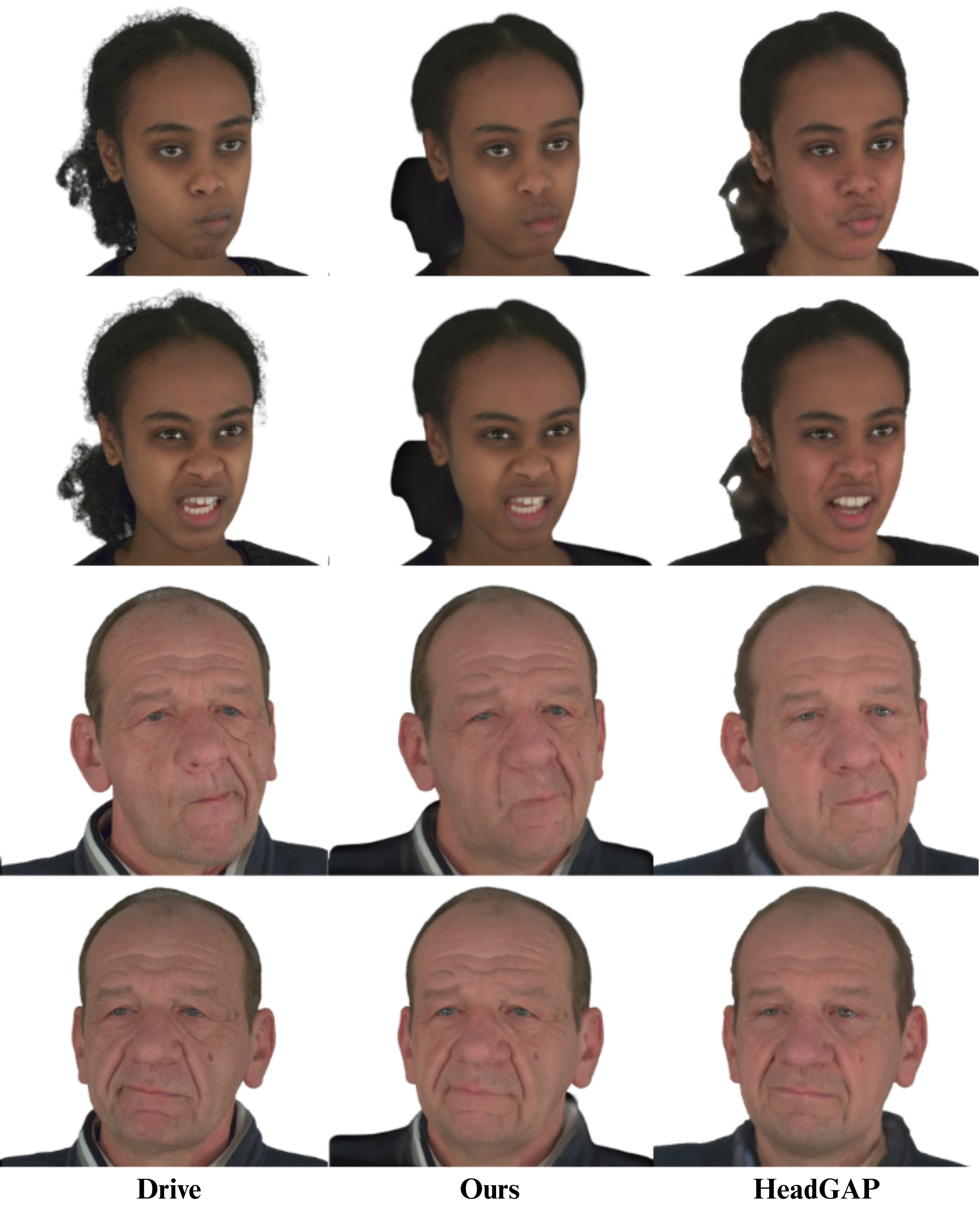}
    \caption{\textcolor{revcolor}{Qualitative comparison against HeadGAP~\cite{zheng2024headgap} under single-image input setting.}}
    \label{fig:compHG}
\end{figure}

As demonstrated in \Cref{tab:cross}, our method achieves superior performance across all evaluation metrics, establishing new state-of-the-art results for cross-identity reenactment. The quantitative results confirm our method's exceptional capability in maintaining identity consistency while accurately transferring complex facial expressions from driving sequences. Notably, our approach outperforms existing methods with strong performance in CSIM (0.8517), AKD (7.8183), and AED (3.2697) metrics. Our method demonstrates clear advantages over all baseline approaches.

These quantitative improvements are particularly significant in the cross-identity setting, where maintaining the source identity while accurately transferring target expressions presents substantial challenges. The superior performance across all three specialized metrics validates our method's effectiveness in disentangling identity and expression features.

% \begin{figure}[!htbp]
%     \centering
%     \includegraphics[width=0.95\linewidth]{figures/headgap.png}
%     \caption{\textcolor{revcolor}{Qualitative comparison against HeadGAP~\cite{zheng2024headgap} under single-image input setting.}}
%     \label{fig:compHG}
% \end{figure}

\begin{figure}[!htbp]
    \centering
    \includegraphics[width=0.85\linewidth]{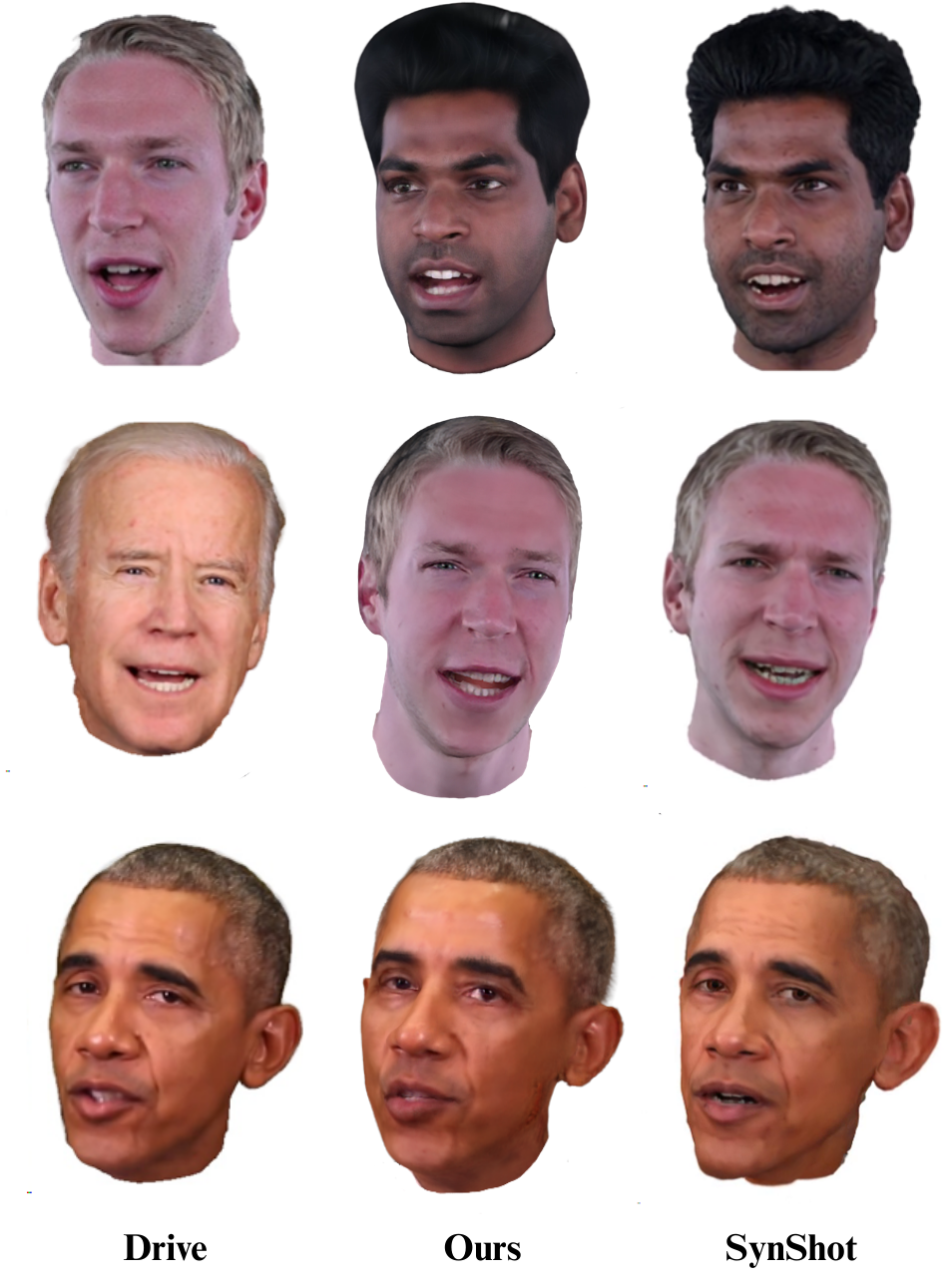}
    \caption{\textcolor{revcolor}{Qualitative comparison with SynShot~\cite{zielonka2025syntheticpriorfewshotdrivable} (few-shot, 3 images) versus our method (single image).}}
    \label{fig:compSynShot}
\end{figure}

\vspace{1mm}
\noindent\textbf{Cross-Identity Reenactment Qualitative Comparison.}
Qualitative comparisons in \Cref{fig:cross} further demonstrate the visual superiority of our approach compared to baseline methods across different data sources. Our method consistently produces more accurate expression transfer while preserving identity-specific features, resulting in higher visual fidelity and more realistic facial animations.

% \begin{figure*}[!ht]
%     \centering
%     \vspace{-2mm}
%     \includegraphics[width=1\textwidth]{figures/cross-comparison6.png}
%     \caption{
%     Cross-identity reenactment qualitative comparison across NeRSemble and in-the-wild data. We compare our method with state-of-the-art baseline approaches including GPAvatar, VOODOO3D, Portrait4D-v1, Portrait4D-v2, GAGAvatar, and LAM. The first four rows show results on NeRSemble dataset subjects with controlled studio conditions, while the bottom three rows demonstrate results on in-the-wild data captured using different smartphone models under varying lighting conditions. Each row shows cross-identity reenactment results where the source identity (leftmost column) is animated using driving expressions from different subjects. Our method consistently produces more accurate expression transfer while better preserving identity-specific features compared to baseline methods across both controlled and challenging real-world scenarios.
% }
%     \label{fig:cross}
%     \vspace{-2mm}
% \end{figure*}

The visual results in \Cref{fig:cross} corroborate our quantitative findings, demonstrating that our method successfully maintains identity characteristics while accurately transferring complex facial expressions across different subjects. Notably, our method shows robust performance on both NeRSemble data (rows 1-4) and challenging in-the-wild data (rows 5-7), where the latter were captured using different smartphone models under varying lighting conditions. The consistent quality across various baseline comparisons and data sources highlights the robustness and generalizability of our approach.

The quantitative superiority stems from our method's effective integration of 2D and 3D priors, which enables robust identity-expression disentanglement. The hierarchical dynamic-static framework ensures that identity-specific features are preserved in static regions while allowing precise expression control in dynamic facial areas. This architectural design, combined with our displacement VAE for fine-grained geometric details, results in high-fidelity cross-identity reenactment that surpasses existing approaches in both objective metrics and visual quality.

\vspace{1mm}
\noindent\textbf{More Comparisons.}
\textcolor{revcolor}{We also compare with HeadGAP~\cite{zheng2024headgap}, which is originally designed as a few-shot method that requires simultaneously captured multi-view inputs to build generalizable 3D head avatars via Gaussian Splatting priors. Since its original implementation is not publicly available, we obtained an updated version from the authors that incorporates architectural improvements and is trained on an expanded dataset. Moreover, HeadGAP applies an additional CNN-based refinement network on top of the Gaussian Splatting rendering output to further enhance image quality. Despite these advantages, under a fair single-image input setting, our method still demonstrates superior performance. As illustrated in \Cref{fig:compHG}, our method consistently achieves better visual quality across different head poses, particularly in terms of identity preservation and expression fidelity. Quantitatively, our approach outperforms HeadGAP in PSNR (24.9656 vs.\ 24.0412) and SSIM (0.8266 vs.\ 0.8251), while HeadGAP achieves a slightly lower LPIPS (0.2164 vs.\ 0.2319).}

% \begin{figure}[!htbp]
%     \centering
%     \includegraphics[width=0.95\linewidth]{figures/headgap.png}
%     \caption{\textcolor{revcolor}{Qualitative comparison against HeadGAP~\cite{zheng2024headgap} under single-image input setting.}}
%     \label{fig:compHG}
% \end{figure}

\vspace{1mm}
\noindent\textcolor{revcolor}{
We also compare with SynShot~\cite{zielonka2025syntheticpriorfewshotdrivable}, a few-shot method that requires 3 input images per identity. Notably, SynShot encodes identity from UV-space texture maps paired with a position map derived from a geometrically accurate, subject-specific FLAME mesh reconstruction. Obtaining such a geometrically accurate per-subject mesh and the corresponding UV texture is a notoriously difficult process. In contrast, our method directly encodes identity from a standard face image using 2D vision priors and only requires standard FLAME parameter estimation to obtain a base position map. Identity-specific geometric details beyond the standard FLAME model are then captured through learned offset maps ($M_\text{offset}$, $M_\text{disp}$). This design lowers the barrier for practical deployment. As shown in \Cref{fig:compSynShot}, our method achieves slightly better visual quality using only a single image as input, whereas SynShot relies on 3 images.}

% \begin{figure}[!htbp]
%     \centering
%     \includegraphics[width=0.85\linewidth]{figures/Syn.png}
%     \caption{\textcolor{revcolor}{Qualitative comparison with SynShot~\cite{zielonka2025syntheticpriorfewshotdrivable} (few-shot, 3 images) versus our method (single image).}}
%     \label{fig:compSynShot}
% \end{figure}

\begin{figure}[!htbp]
    \centering
    \includegraphics[width=\linewidth]{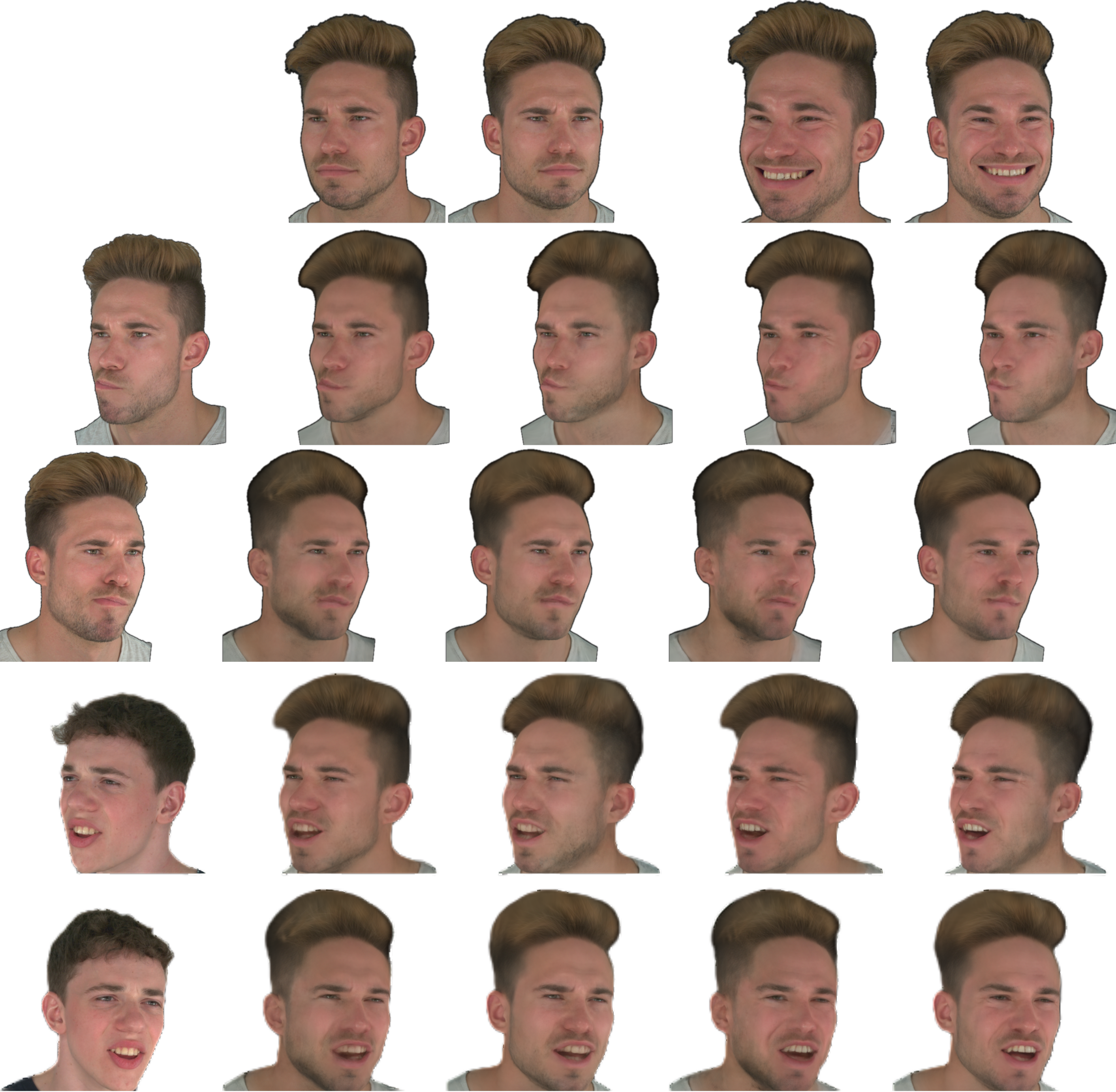}
    \caption{\textcolor{revcolor}{Multi-view and multi-expression analysis. Row 1: four source images (2 expressions $\times$ 2 side views). Rows 2--3: self-reenactment. Rows 4--5: cross-identity reenactment. Results remain consistent across all conditions.}}
    \label{fig:viewanalysis}
\end{figure}

\begin{figure}[t]
    \centering
    \includegraphics[width=\linewidth]{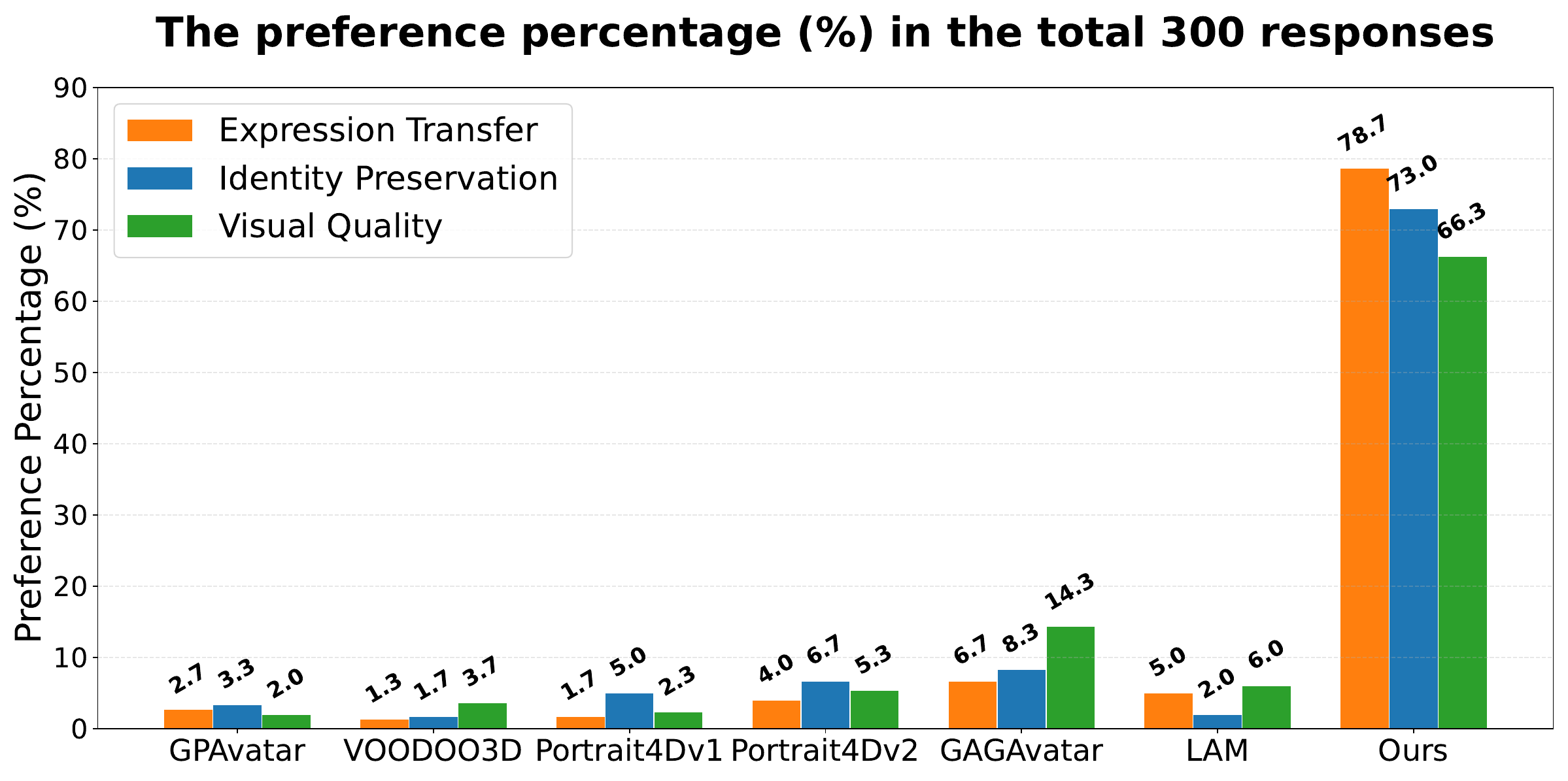}
    \caption{\textcolor{revcolor}{User study results on cross-identity reenactment. Our method achieves the highest preference rate across all three evaluation criteria: identity preservation, expression similarity, and visual quality.}}
    \label{fig:userstudy}
\end{figure}

\vspace{1mm}
\noindent\textcolor{revcolor}{\textbf{User Study.}
To further evaluate cross-identity reenactment quality beyond automatic metrics, we conduct a user study with 60 participants. We build a question bank of 7 cross-reenactment comparisons, each displaying the source image, driving image, and results from 7 methods (ours and 6 baselines) in randomized order. Each participant is randomly presented with 5 out of 7 comparisons and asked three questions per comparison: (1) which result best preserves the source identity, (2) which result best matches the driving expression, and (3) which result has the best overall visual quality. In total we collect 300 responses. As shown in \Cref{fig:userstudy}, our method receives the highest preference rate across all three criteria, confirming its superiority in identity preservation, expression transfer accuracy, and visual quality for cross-identity reenactment. Detailed questionnaire design is provided in the supplementary material.}

\vspace{1mm}
\noindent\textcolor{revcolor}{\textbf{Multi-View and Multi-Expression Analysis.}
We conduct a comprehensive analysis of our method across diverse source views and expressions. As shown in \Cref{fig:viewanalysis}, we construct a source set of 4 images from the same identity with two expressions (neutral and smiling) each captured from two side viewpoints (left and right). Rows 2--3 show self-reenactment results where the same expression is driven across different rendered views, while rows 4--5 present cross-identity reenactment under the same setting. As reported in \Cref{tab:viewanalysis}, our method maintains stable performance across all four input conditions. The Front-Smile configuration achieves the best results in most metrics (e.g., PSNR 25.85, LPIPS 0.1836, AKD 6.50), which is expected since it provides the richest facial detail as input. Nevertheless, the other configurations remain competitive---for self-reenactment, PSNR ranges from 24.54 to 25.85 and SSIM stays within 0.84--0.85; for cross-reenactment, CSIM remains consistently above 0.95 across all conditions. These results confirm that SEGA generalizes well across varying input poses and expressions without notable performance degradation.}

\textcolor{revcolor}{\begin{table*}[!htbp]
  \centering
  \begin{tabularx}{\textwidth}{l*{6}{X}|*{3}{X}}
    \toprule
    \multirow{2}{*}{Source Input} & \multicolumn{6}{c|}{Self-Reenactment} & \multicolumn{3}{c}{Cross-Reenactment} \\
    \cmidrule(lr){2-7} \cmidrule(lr){8-10}
     & PSNR$\uparrow$ & SSIM$\uparrow$ & LPIPS$\downarrow$ & CSIM$\uparrow$ & AKD$\downarrow$ & AED$\downarrow$ & CSIM$\uparrow$ & AKD$\downarrow$ & AED$\downarrow$ \\
    \midrule
    Side-Neutral  & 24.9534 & 0.8499 & 0.1841 & 0.8783 & 6.5077 & 2.6872 & 0.9666 & 7.5523 & 3.5766 \\
    Front-Neutral & 24.5360 & 0.8439 & 0.1932 & 0.8738 & 6.7025 & 2.6718 & 0.9625 & 7.6368 & 4.1708 \\
    Side-Smile    & 25.2123 & 0.8503 & 0.1909 & 0.8756 & 6.6644 & 2.7371 & 0.9544 & 7.4025 & 3.7248 \\
    \rowcolor[rgb]{.906,.902,.902} Front-Smile   & \textbf{25.8480} & \textbf{0.8506} & \textbf{0.1836} & \textbf{0.8795} & \textbf{6.5005} & \textbf{2.5830} & \textbf{0.9692} & \textbf{7.2795} & \textbf{3.3385} \\
    \bottomrule
  \end{tabularx}%
  \caption{\textcolor{revcolor}{Quantitative analysis across different source input conditions for self-reenactment and cross-identity reenactment.}}
  \label{tab:viewanalysis}%
\end{table*}}

\begin{figure}[!htbp]
    \centering
    \includegraphics[width=\linewidth]{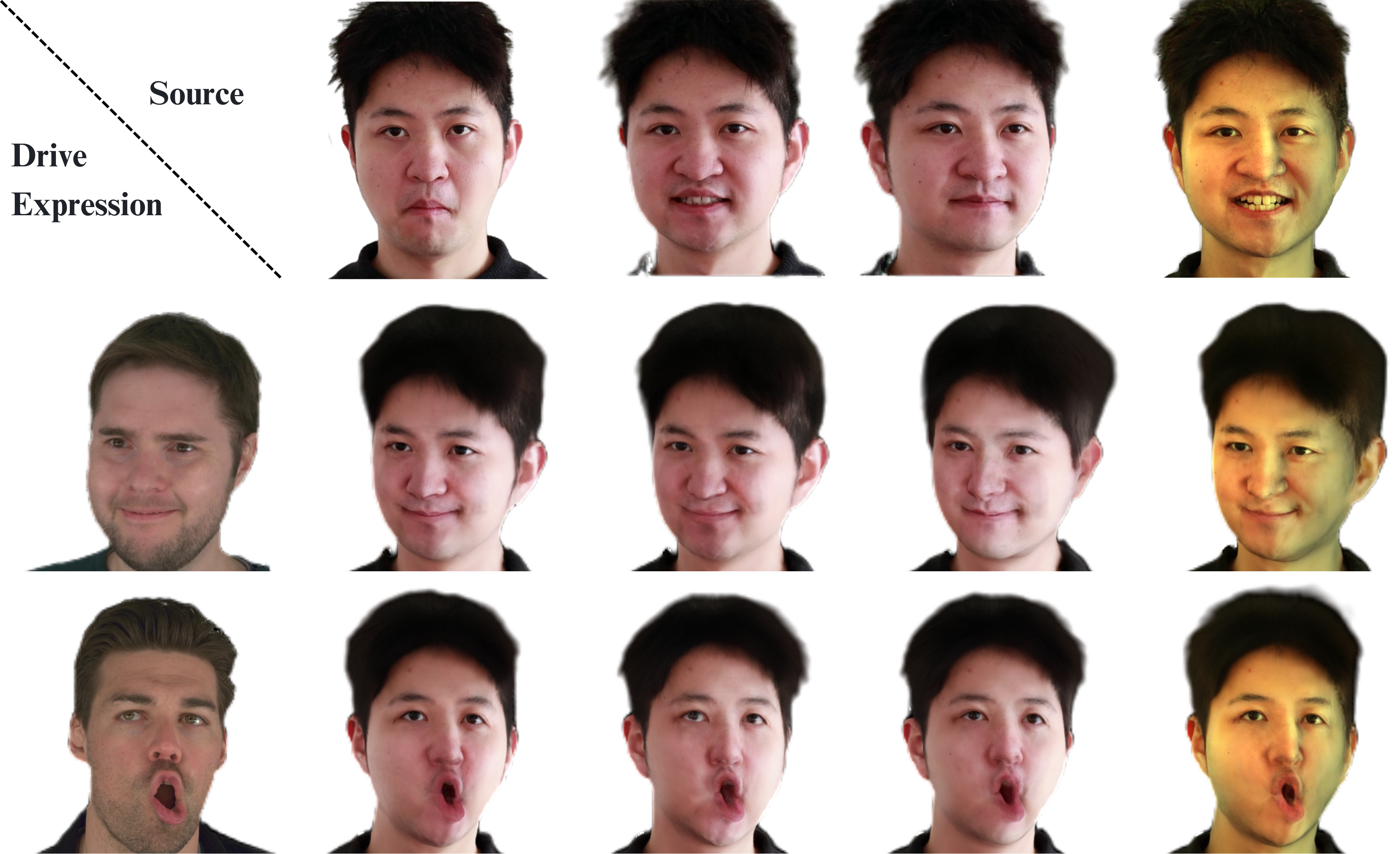}
    \caption{\textcolor{revcolor}{Reenactment from in-the-wild inputs with varying poses, expressions, and lighting (including synthetic relighting). Our method produces consistent results across all conditions.}}
    \label{fig:itwcompare}
\end{figure}

\vspace{1mm}
\noindent\textcolor{revcolor}{\textbf{Robustness to In-the-Wild Lighting Conditions.}
To evaluate our method beyond controlled studio settings, we test on in-the-wild images captured under challenging illumination. As shown in \Cref{fig:itwcompare}, the four source columns depict the same identity: the first three are captured under high-contrast directional lighting with varying expressions, and the fourth is synthetically re-lit using ClipDrop Relight~\cite{clipdrop_relight} to simulate an extreme, uncommon lighting condition. Each source is driven by two different target expressions (rows 2--3). Despite the significant variation in lighting and expression across inputs, our method consistently produces faithful reenactment results that preserve both identity and expression, demonstrating strong robustness to diverse real-world illumination conditions.}

\begin{figure}[t]
    \centering
    \includegraphics[width=0.5\textwidth]{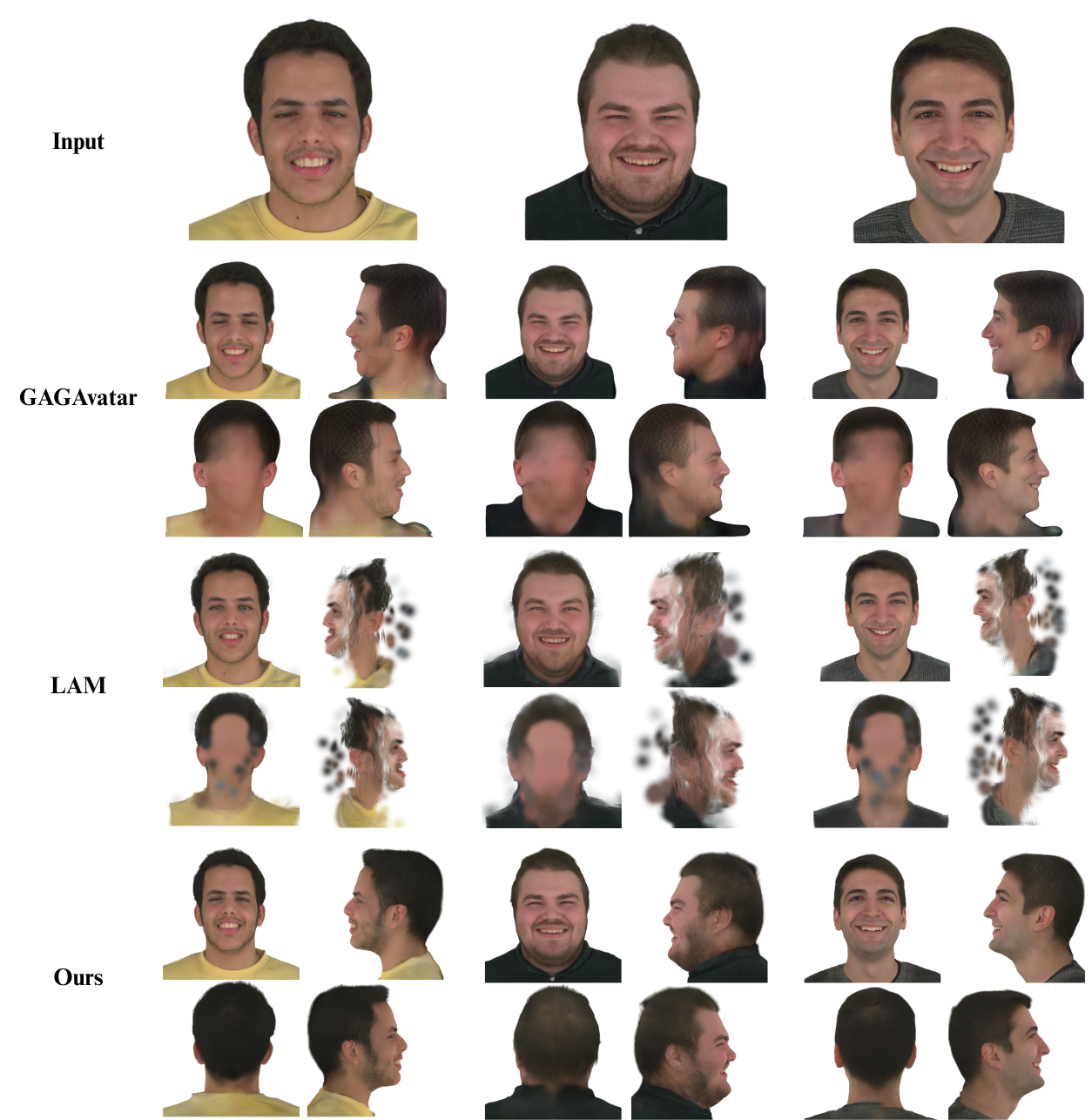}
    \caption{Novel view synthesis results showing multi-view consistency. Our method generates high-quality renderings from different viewpoints around the Y-axis at  0°, -90°, 90°and 180°, demonstrating robust  view-consistent facial details without artifacts.}
    \label{fig:NovelView}
    \vspace{-2mm}
\end{figure}

\vspace{1mm}
\noindent\textbf{Novel View Synthesis.}
To evaluate the 3D consistency and geometric fidelity of our method, we conduct comprehensive novel view synthesis experiments. For fair comparison, we focus on methods that also use Gaussian-based representations, specifically GAGAvatar~\cite{chu2024generalizable} and LAM~\cite{he2025LAM}, as they share similar rendering paradigms with our approach. As illustrated in \Cref{fig:NovelView}, we render avatars from multiple viewpoints by rotating around the Y-axis at 0°, 90°, -90°, and 180°, demonstrating our method's capability to generate photorealistic renderings from previously unseen camera angles. The results showcase remarkable multi-view consistency across different viewing angles, with no visible artifacts, geometric distortions, or unrealistic facial expressions.

Our hierarchical UV-space Gaussian Splatting framework effectively maintains spatial coherence, ensuring that facial features and fine details like teeth remain geometrically consistent across all viewpoints. The consistent lighting, shading, and texture details across different angles demonstrate that our approach successfully captures the underlying 3D geometry rather than merely interpolating between 2D views, validating the effectiveness of our single-image-based 3D avatar creation pipeline. Compared to GAGAvatar and LAM, our method shows superior geometric consistency and detail preservation across novel viewpoints, particularly in challenging regions like the mouth and eye areas.
% \input{Tabs/quantitative}

% \begin{figure*}[htb]  % 添加更多位置选项
%     \centering
%     % \vspace{-1.5mm}
%     \includegraphics[width=\textwidth]{figures/novelview6.png}
%     \caption{
%         Novel view synthesis results showing multi-view consistency. Our method generates high-quality renderings from different viewpoints around the Y-axis at  0°, -90°, 90°and 180°, demonstrating robust  view-consistent facial details without artifacts.
%     }
%     \label{fig:NovelView}
%     % \vspace{-1.5mm}
% \end{figure*}

\subsection{Ablation Study}
% \begin{table*}[htbp]
%   \centering
%   \caption{Quantitative ablation study comparing the performance of different configurations.}
%   \begin{tabularx}{\textwidth}{l*{8}{X}}
%     % \begin{tabular}{ccccccccc}
%     \toprule
%     \multirow{2}[3]{*}{Method} & \multicolumn{3}{c}{Frontal view} & \multicolumn{3}{c}{All views} \\
% \cmidrule{2-7}          & PSNR  & SSIM  & \multicolumn{1}{l}{LPIPS}  & PSNR  & SSIM  & \multicolumn{1}{l}{LPIPS}  \\
%     \midrule
%     Full-static variation &       &       &       &        &       &  \\
%     Full-dynamic variation &       &       &       &          &       &  \\
    
%     \midrule
%     Baseline method  &       &       &       &       &         &  \\
%     - Identity loss  &       &       &       &       &        &  \\
%     \quad - Perceptual loss  &       &       &       &            &       &  \\
%     - 2D prior &       &       &       &       &         &  \\
%     \rowcolor[rgb]{ .906,  .902,  .902} + Finetuning (Full method)  &       &          &       &       &       &  \\
%     % w/o \cite{zheng2024headgap} &       &       &       &       &       &       &       &  \\
%     \bottomrule
    
%     % \end{tabular}%
%     \end{tabularx}%
%   \label{tab:ablation}%
% \end{table*}%

\begin{table}[!htbp]
  \centering

  \begin{tabularx}{0.45\textwidth}{l*{4}{X}}
    % \begin{tabular}{ccccccccc}
    \toprule
%     \multirow{2}[3]{*}{Method} & \multicolumn{3}{c}{Frontal view} & \multicolumn{3}{c}{All views} \\
% \cmidrule{2-7}          & PSNR  & SSIM  & \multicolumn{1}{l}{LPIPS}  & PSNR  & SSIM  & \multicolumn{1}{l}{LPIPS} 

    Method &    PSNR$\uparrow$  & SSIM$\uparrow$  &LPIPS$\downarrow$  \\
    \midrule
    Full-static &  23.7895     &  0.8012     & 0.2498     \\
    Full-dynamic & 24.9364      & 0.8296      &  0.2355     \\
    
    \midrule
    w/o VGG loss & 24.8365      &  0.8455    & 0.2398      \\
    w/o 2D Prior & 25.1398      &  0.8604           & 0.2202 \\
    w/o Finetune  &  23.7512     &   0.8316        & 0.2457 \\
    \rowcolor[rgb]{ .906,  .902,  .902} Ours  &  \textbf{\textbf{25.3311}}      &  \textbf{0.8638}     & \textbf{0.2187} \\
    % w/o \cite{zheng2024headgap} &       &       &       &       &       &       &       &  \\
    \bottomrule
    
    % \end{tabular}%
    \end{tabularx}%
      \caption{Quantitative Ablation Study.}
  \label{tab:ablation}%
\end{table}%

\begin{figure}[htbp!]
      \centering
      \includegraphics[width=\columnwidth]{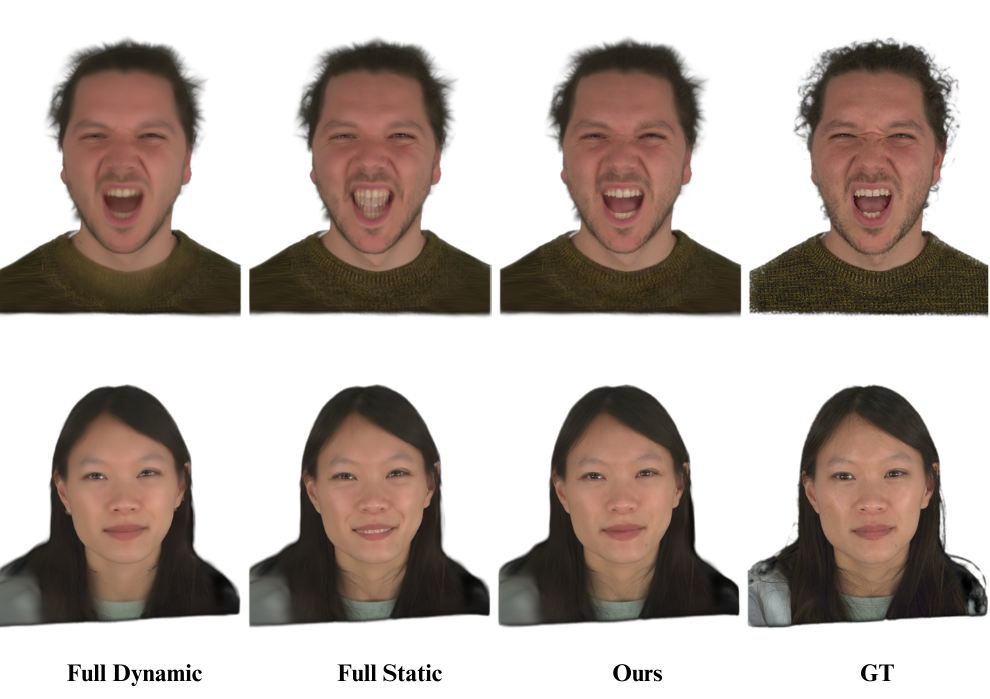}
      % \vspace{-1.5mm}
      \caption{
      Ablation study evaluating different configurations of our hierarchical dynamic-static framework. We compare the full method against variants using only static or dynamic branches to demonstrate the superior performance of our combined approach.
      }
      % \vspace{-5mm}
      \label{fig:ablation1}
  \end{figure}
\begin{figure}[htbp!]
      \centering
      \includegraphics[width=\columnwidth]{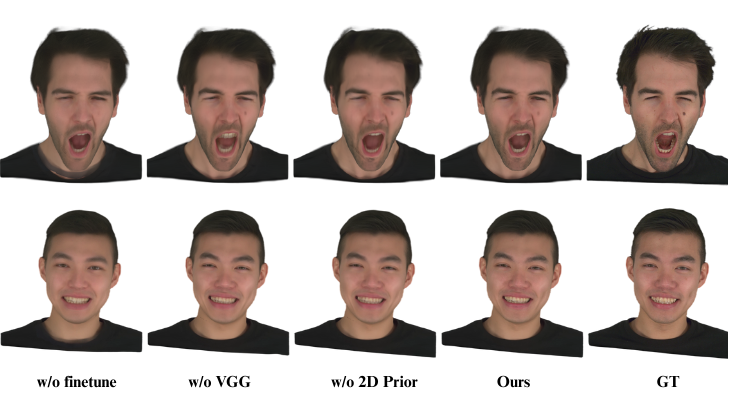}
      % \vspace{-1.5mm}
      \caption{
      Ablation study evaluating the impact of different components including 2D prior integration, loss functions, and personalized fine-tuning. The results demonstrate how each component contributes to the final rendering quality.
      }
      % \vspace{-5mm}
      \label{fig:ablation2}
  \end{figure}

In this subsection, we conduct ablation studies to evaluate our key components. The qualitative results on the ``031, 042, 286, 290'' ID from NeRSemble~\cite{kirschstein2023nersemble} are shown in \Cref{fig:ablation1} and \Cref{fig:ablation2}.
% , where the baseline model does not use finetuning, and comparisons before finetuning help isolate the effects of other components, providing a clearer view of generalization and performance. 

\noindent\textbf{Hierarchical Dynamic-Static Framework.}
We evaluate our hierarchical dynamic-static framework with two variations: (1) full-static, using only the static branch, and (2) full-dynamic, relying solely on the dynamic branch to generate Gaussian parameters on a $1024 \times 1024$ UV map. As shown in \Cref{tab:ablation} and \Cref{fig:ablation1}, integrating both priors yields the best performance across all metrics while preserving facial details. Moreover, it reduces computation time from \textbf{240ms} (dynamic) to \textbf{50ms} by generating Gaussian parameters on a smaller UV map during inference.

\noindent\textbf{2D Prior Integration.}
We assess the 2D prior by comparing model without encoder pretraining. Training from scratch reduces generalization, especially in local facial details (\Cref{fig:ablation2}), underscoring the 2D prior's importance.

% We assess the role of the 2D prior by comparing the model with and without pretraining the encoder.
% Removing the 2D prior (training from scratch) results in decreased generalization, particularly in preserving local facial details, as shown \Cref{fig:ablation}, highlighting the 2D prior's importance in improving generalization.

\noindent\textbf{Loss Function Analysis.}
We assess loss function impact by excluding the perceptual VGG loss. While numerical differences in \Cref{tab:ablation} seem minor, \Cref{fig:ablation2} shows that perceptual loss is crucial for fine details (e.g., mouth).

\noindent\textbf{Personalized Finetuning.}
As shown in \Cref{tab:ablation} and \Cref{fig:ablation2}, personalized finetuning significantly enhances facial reconstruction. Without finetuning, avatars resemble the input less, while the full method refines details and expressions for a closer match.

\section{Conclusion}
\label{sec:conclusion}

% We introduced SEGA, a novel approach for creating photorealistic 3D head avatars from a single image. By combining generalized priors derived from large-scale 2D datasets with 3D priors learned from multi-view data, SEGA achieves robust generalization to unseen identities while ensuring 3D consistency across novel viewpoints and expressions. Our hierarchical UV-space Gaussian Splatting framework efficiently separates dynamic and static facial components through a dual-branch architecture, with the dynamic branch encoding expression-driven details and the static branch focusing on expression-invariant regions.

We propose SEGA, a novel method for creating high-quality, fully 360-degree renderable head avatars from a single image.
Our approach bridges 2D identity diversity and 3D geometric consistency through two key insights.
First, a hierarchical static--dynamic decomposition enables specialized processing of rigid and deformable facial regions with real-time performance.
Second, we strategically integrate 2D priors (DINOv2 and VQ-VAE encoders) with 3D data via joint training and displacement VAE refinement.
Experimental validation demonstrates that SEGA outperforms existing methods across generalization to novel views, robust expression animation, and high identity diversity, making it particularly suitable for practical applications in virtual reality, telepresence, and digital entertainment.

% We propose SEGA, a novel approach for creating 3D head avatars from a single image, addressing the limitations of multi-view methods. By combining 2D priors (DINOv2 pretrained on large-scale 2D data for general visual features, CodeFormer pretrained on large-scale 2D face data for face-specific features) with extensive 3D multi-view data through a hierarchical UV-space Gaussian Splatting framework, SEGA achieves robust generalization and high-quality avatar personalization. Our dual dynamic-static decomposition preserves geometric consistency and facial details while achieving efficient inference (50ms per frame). Crucially, person-specific fine-tuning further enhances avatar fidelity. Experimental results show that SEGA outperforms state-of-the-art methods in generalization, identity preservation, and expression realism, paving the way for broader one-shot avatar creation in practical applications like virtual reality and digital entertainment.

\noindent\textbf{Limitations and Future Work.}
Our method still faces some limitations. First, it struggles with avatars of subjects wearing glasses or facial accessories because the training data lacks these samples. 
Second, we cannot model non-rigid dynamic hair movements, as our framework focuses on facial regions with relatively stable hair geometry.
%
% Future work will address these issues by incorporating diverse training data and improving dynamic hair modeling for more comprehensive avatar rendering.
Future work will address these issues by incorporating more diverse training data and introducing a dedicated hair modeling module to for more comprehensive avatar rendering.

\noindent\textbf{Broader Impact.}
While our technology democratizes avatar creation for VR, telepresence, and entertainment, we acknowledge potential misuse risks. Our method produces detectable artifacts and requires technical expertise, creating natural barriers to abuse. We advocate for robust detection mechanisms and ethical guidelines to ensure responsible deployment.

\bibliographystyle{IEEEtran}

% \bibitem{ref1}
% {\it{Mathematics Into Type}}. American Mathematical Society. [Online]. Available: https://www.ams.org/arc/styleguide/mit-2.pdf

% \bibitem{ref2}
% T. W. Chaundy, P. R. Barrett and C. Batey, {\it{The Printing of Mathematics}}. London, U.K., Oxford Univ. Press, 1954.

% \bibitem{ref3}
% F. Mittelbach and M. Goossens, {\it{The \LaTeX Companion}}, 2nd ed. Boston, MA, USA: Pearson, 2004.

% \bibitem{ref4}
% G. Gr\"atzer, {\it{More Math Into LaTeX}}, New York, NY, USA: Springer, 2007.

% \bibitem{ref5}M. Letourneau and J. W. Sharp, {\it{AMS-StyleGuide-online.pdf,}} American Mathematical Society, Providence, RI, USA, [Online]. Available: http://www.ams.org/arc/styleguide/index.html

% \bibitem{ref6}
% H. Sira-Ramirez, ``On the sliding mode control of nonlinear systems,'' \textit{Syst. Control Lett.}, vol. 19, pp. 303--312, 1992.

% \bibitem{ref7}
% A. Levant, ``Exact differentiation of signals with unbounded higher derivatives,''  in \textit{Proc. 45th IEEE Conf. Decis.
% Control}, San Diego, CA, USA, 2006, pp. 5585--5590. DOI: 10.1109/CDC.2006.377165.

% \bibitem{ref8}
% M. Fliess, C. Join, and H. Sira-Ramirez, ``Non-linear estimation is easy,'' \textit{Int. J. Model., Ident. Control}, vol. 4, no. 1, pp. 12--27, 2008.

% \bibitem{ref9}
% R. Ortega, A. Astolfi, G. Bastin, and H. Rodriguez, ``Stabilization of food-chain systems using a port-controlled Hamiltonian description,'' in \textit{Proc. Amer. Control Conf.}, Chicago, IL, USA,
% 2000, pp. 2245--2249.
\bibliography{template}
% \end{thebibliography}

% \newpage

 

% \end{IEEEbiography}



\vfill

\end{document}